\documentclass[reprint,superscriptaddress,amsmath,amssymb,aps,nolongbibliography]{revtex4-2}

\usepackage{mathtools}
\usepackage{graphicx}
\usepackage[colorlinks=true,linkcolor=blue,citecolor=blue,urlcolor=blue,pdfauthor={ },pdftitle={ },pdfsubject={ },pdfkeywords={ }]{hyperref}
\usepackage{times}
\usepackage[qm]{qcircuit}
\usepackage{braket}
\usepackage{enumitem}
\usepackage{dcolumn}
\usepackage{bm}

\usepackage{siunitx}
\usepackage{upgreek}

\usepackage{color}
\usepackage{tikz}
\usepackage{pgf}
\usetikzlibrary{spy}

\providecommand{\openone}{\leavevmode\hbox{\small1\kern-3.8pt\normalsize1}}

\newcommand{\map}[1]{\mathcal{#1}}

\def\>{\rangle}
\def\<{\langle}

\makeatletter

\newcommand{\Rmnum}[1]{\expandafter\@slowromancap\romannumeral #1@}
\makeatother

\begin{document}
	
\title{Anomalous heat flow and quantum Otto cycle with indefinite causal order}
\author{Qing-Feng Xue}
\author{Qi Zhang}
\author{Xu-Cai Zhuang}
\author{Yun-Jie Xia}
\affiliation{School of Physics and Physical Engineering, Qufu Normal University, 273165, Qufu, China}

\author{Enrico Russo}
\affiliation{Dipartimento di Ingegneria, Università degli Studi di Palermo, Viale delle Scienze, 90128 Palermo, Italy}
\affiliation{University San Pablo-CEU, CEU Universities, Department of Applied Mathematics and Data Science, Campus de Moncloa, C/Juli\'{a}n Romea 23 28003, Madrid, Spain}

\author{Giulio Chiribella}
\email{giulio@cs.hku.hk}
\affiliation{QICI Quantum Information and Computation Initiative, School of Computing and Data  Science, The University of Hong Kong, Pokfulam Road, Hong Kong}
\affiliation{Department of Computer Science, University of Oxford, Wolfson Building, Parks Road, Oxford, UK}
\affiliation{Perimeter Institute for Theoretical Physics, 31 Caroline Street North, Waterloo,  Ontario, Canada}

\author{Rosario Lo Franco}
\email{rosario.lofranco@unipa.it}
\affiliation{Dipartimento di Ingegneria, Università degli Studi di Palermo, Viale delle Scienze, 90128 Palermo, Italy}

\author{Zhong-Xiao Man}
\email{zxman@qfnu.edu.cn}
\affiliation{School of Physics and Physical Engineering, Qufu Normal University, 273165, Qufu, China}	

\begin{abstract}
The principle that heat spontaneously flows from higher temperature to lower temperature
is a cornerstone of classical thermodynamics, often assumed to be independent of the sequence of interactions.
While this holds true for macroscopic systems at equilibrium, here we show that,
when the order of interactions between two identical thermalization channels is indefinite,
an anomalous heat flow emerges, whereby heat can sometime flow from a colder entity to a hotter one.
Taking advantage of this anomalous heat flow, we design a quantum Otto cycle with indefinite causal order, which not only achieves refrigeration but also generates work.   The anomalous heat flow and the quantum Otto cycle are experimentally simulated in a photonic quantum setup, which provides a proof-of-principle demonstration of the theory. 	

\end{abstract}

\maketitle

\emph{\bfseries Introduction.}---In classical physics, 
causality is fundamentally well-defined: given two events $A$ and $B$ of sufficiently short duration,
there are only three possibilities: either $A$ causally preceeds $B$, or $B$ causally preceeds $A$,
or $A$ and $B$ are spacelike. 
However, this reassuring situation no longer holds in quantum mechanics,
which is in principle compatible with situations where the order of two events is indefinite \cite{chiribella2009beyond,ICO1,ICOappli9QS}.
For example, the order in which events $A$ and $B$ take
place could be controlled by the state of a qubit, generating a coherent superposition
of the scenario in which $A$ occurs before $B$ and the scenario in which $B$ occurs 
before $A$, in a process known as the quantum switch \cite{chiribella2009beyond,ICOappli9QS}.
Indefinite causal order (ICO) challenges our understanding of causality and provides
insights in the operational features of a future theory of quantum gravity
\cite{Hardy2007,Spekkens07,Giacomini2019,Giacomini2021spacetimequantum,Castro-Ruiz2020}.
In addition, it has been shown to offer advantages in a wide range of quantum information tasks,
including quantum channel discrimination \cite{ICOappli1,ICOappli16},
communication complexity \cite{ICOappli2,ICOappli3}, 
quantum computation \cite{ICOappli9QS,ICOappli10,ICOappli11,ICOappli12}, 
quantum metrology \cite{ICOappli13,ICOappli14}, and inversion of unitary gates \cite{ICOappli15}.
These advantages have spurred a series of experiments,
particularly in photonic systems, whose setups were inspired by the quantum switch
\cite{ICOMach4,ICObatt,ICOexphoto0,ICOexphoto1,ICOexphoto2,ICOexphoto3,ICOexphoto4,ICOexphoto5,ICOexphoto6,ICOexphoto7}.

The growing interest in ICO stimulated an exploration of new communication scenarios, 
in which the transmitted information undergoes multiple noisy processes acting in an indefinite order  \cite{ICOappli5,ICOappli6,ICOappli7,ICOappli8}. In turn, this research line motivated a new research direction in quantum thermodynamics, thanks to the connection between information and entropy. In particular, a series of works studied the implication of ICO for work extraction \cite{ICOWork1,ICOWork2,ICOWork3,simonov2025activation}, 
to drive the operation of quantum heat machines \cite{ICOMach1,ICOMach2,ICOMach3,ICOMach4}
and to charge quantum batteries \cite{ICObatt}.
Applying ICO to thermodynamics not only extends the traditional
thermodynamic paradigms but also reveals new features of ICO, which emerges as a counterintuitive thermodynamic resource \cite{ICOMach1,ICOMach2,ICOMach3,ICOMach4,ICObatt}.

In this work, we demonstrate that ICO can induce anomalous heat transfers, 
which can be harnessed to perform thermodynamic tasks.
We show that even if the temperature of the system is higher than the
thermalization temperature of the channels, heat can flow into the system, and conversely.
For this to happen,
the temperature difference
between the system and the channels needs to stay below
a certain threshold, and it also depends on the measurement outcomes
of the control system.
We further show that 
this anomalous heat flow can be exploited in an ICO Otto cycle which transfers heat from a low-temperature environment to a high-temperature one while performing work.  
The anomalous heat flow and ICO Otto cycle are simulated experimentally
in a proof-of-principle demonstration using a photonic setup.

\begin{figure}[htbp]
	\begin{center}
	{\includegraphics[width=0.9\linewidth]{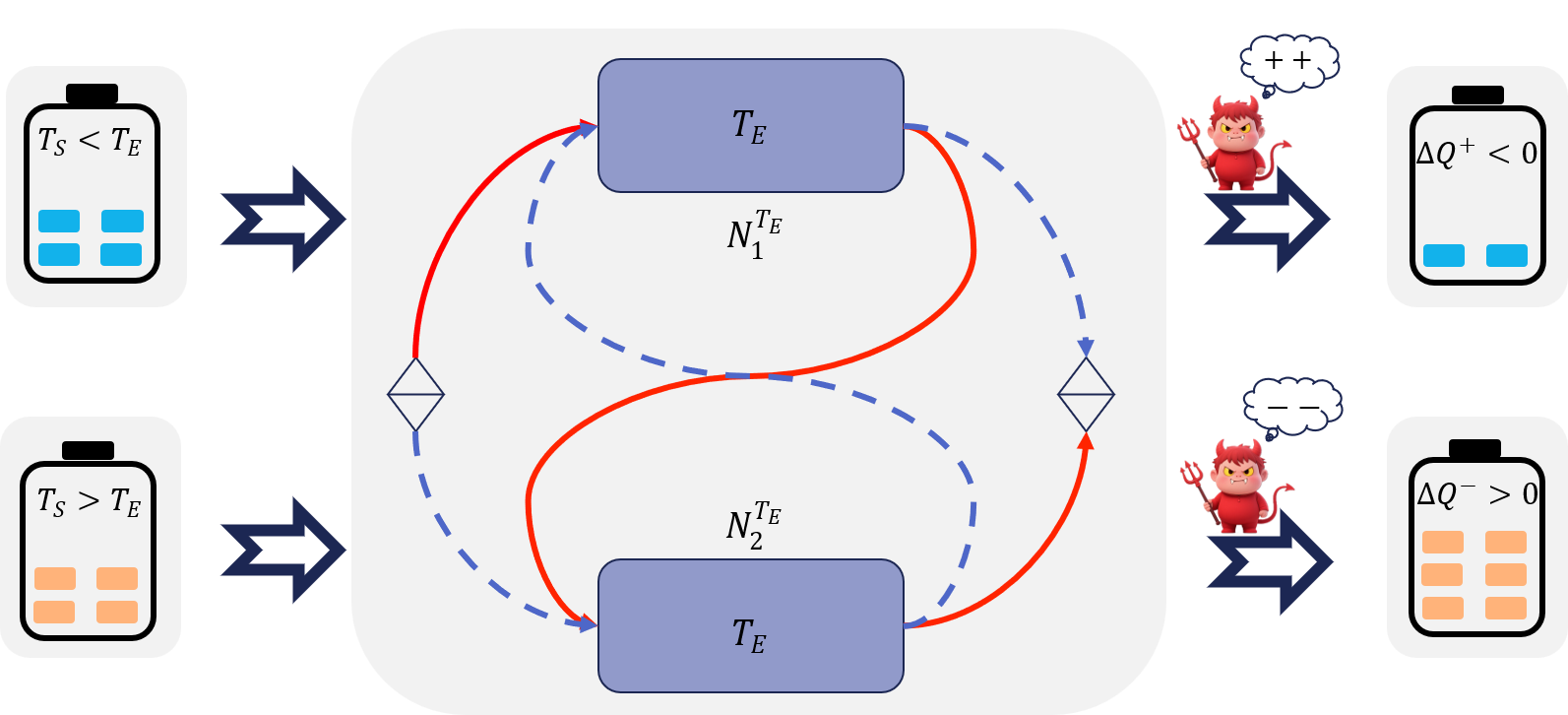} }
	\end{center}
	\caption{Illustration of heat exchanges between the system 
(initially at temperature $T_{\texttt{S}}$, represented by a bottle with broken lines indicating the heat)
and two identical thermalizing channels, $\mathcal{N}^{T_{\texttt{E}}}_{1}$ and $\mathcal{N}^{T_{\texttt{E}}}_{2}$,
both at temperature $T_{\texttt{E}}$. We analyze two scenarios: $T_{\texttt{S}}<T_{\texttt{E}}$ (upper) and $T_{\texttt{S}}>T_{\texttt{E}}$ (lower). In both cases, the system interacts with an ICO by applying the quantum switch. 
The control qubit is initially prepared in a superposition state and subsequently measured in the
basis $\left\{\left|+\right\rangle_{c},\left|-\right\rangle_{c}\right\}$ after the interaction. 
Anomalous heat flow occurs conditioned on the measurement outcomes of the control qubit:
for $T_{\texttt{S}}<T_{\texttt{E}}$, if $\left|+\right\rangle_{c}$ is detected,
heat is transferred from the cold system to the hot channels, whereas 
for $T_{\texttt{S}}>T_{\texttt{E}}$, if $\left|-\right\rangle_{c}$ is measured,
heat flows from the cold channels to the hot system.   }
	\label{mod}
\end{figure}

\emph{\bfseries Anomalous heat flow driven by ICO.}---
We examine the heat flow in a thermodynamic process where a system $\texttt{S}$
is subjected to two thermalizing channels.
A thermalizing channel $\mathcal{N}^{T_{\texttt{E}}}$ at temperature $T_{\texttt{E}}$ 
transforms any initial state $\rho_{\texttt{S}}^{0}$ of the system 
into a thermal state $\rho_{\texttt{S}}^{T_{\texttt{E}}}$ at temperature $T_{\texttt{E}}$.
When two identical thermalizing channels, $\mathcal{N}^{T_{\texttt{E}}}_{1}$ and $\mathcal{N}^{T_{\texttt{E}}}_{2}$,
both at the same temperature $T_{\texttt{E}}$, act sequentially on the system, the
final state remains $\rho_{\texttt{S}}^{T_{\texttt{E}}}$, 
independent of the order of application.
However, some counter-intuitive results emerge
when the two channels act on the system through an ICO.
This can be achieved by introducing a control qubit that dictates the order in which the channels are applied:
specifically, the channels are applied in the order
$\mathcal{N}^{T_{\texttt{E}}}_{1}\circ\mathcal{N}^{T_{\texttt{E}}}_{2}$
when the control qubit is in state $\left|0\right\rangle_{c}$,
and $\mathcal{N}^{T_{\texttt{E}}}_{2}\circ\mathcal{N}^{T_{\texttt{E}}}_{1}$
if in $\left|1\right\rangle_{c}$.
After the joint system-channels interaction, measuring the control qubit in the basis $\{\left|+\right\rangle_c, \left|-\right\rangle_c\}$, with $\left|\pm\right\rangle_c=\frac{1}{\sqrt{2}}(\left|0\right\rangle_{c}
\pm\left|1\right\rangle_{c})$, 
projects the system into one of two conditional states $\rho_{\texttt{S}}^{\pm}$,
depending on the measurement outcome
(see the Supplementary Material for further details).
Notably, the effective temperatures of the resulting system states vary with the measurement outcomes of the control qubit.
This feature allows us to exploit ICO to control the system’s temperature and achieve a variety of intriguing thermodynamic tasks.

Here, we focus on the heat exchange in the ICO process, defined as
$\Delta Q^{\pm}=P_{\pm}\mathrm{Tr} \left[\left(\rho_{\texttt{S}}^{\pm}-\rho_{\texttt{S}}^{0}\right)H_{\texttt{S}}
\right]$, with $P_{\pm}$ the measurement probabilities and $H_{\texttt{S}}$ the system's Hamiltonian.
The equal system-channel temperatures case was analysed in \cite{ICOMach1}, and tested in \cite{ICOMach3,ICOMach4}. 
Our main observation is that, conditionally on the outcome of a 
measurement on the control qubit, it is possible for the system
to be cooled (heated) even when $T_{\texttt{S}}<T_{\texttt{E}}$ ($T_{\texttt{S}}>T_{\texttt{E}}$) thanks to the ICO (see Fig. \ref{mod}).
There is an anomalous heat flow between the system and channels,
with heat transferring from the lower-temperature entity to the high-temperature one.
This phenomenon occurs if the temperature gradient between system and channels
remains under a certain threshold.
For a system considered as a qubit with Hamiltonian 
$H_S=\frac{\omega_\texttt{S}}{2} \sigma_z$,
and the control qubit prepared in the $\ket{+}_{c}$ state, 
we obtain the following conditions (see Supplemental Material for the derivation):

(i) when $T_{\mathrm{S}} > T_{\mathrm{E}}$, heating of the system happens if 
\begin{equation}\label{con1} 
	T_{\mathrm{S}} < 2 T_{\mathrm{E}};
\end{equation} 

(ii) when $T_{\mathrm{S}} < T_{\mathrm{E}}$, the system is cooled if 
\begin{equation}\label{con2}
	T_{\texttt{E}}<\frac{\omega_{\texttt{S}}}{ 2 \operatorname{artanh} \left( \sinh\left( \frac{\omega_{\texttt{S}}}{T_{\texttt{S}}} \right)/
		\left(\cosh\left(\frac{\omega_{\texttt{S}}}{T_{\texttt{S}}}\right)+2\right) \right)}.
\end{equation}

\emph{\bfseries Physical origin of the anomalous heat flow.}---
\begin{figure}[t]
	\begin{center}
		{\includegraphics[width=0.6\linewidth]{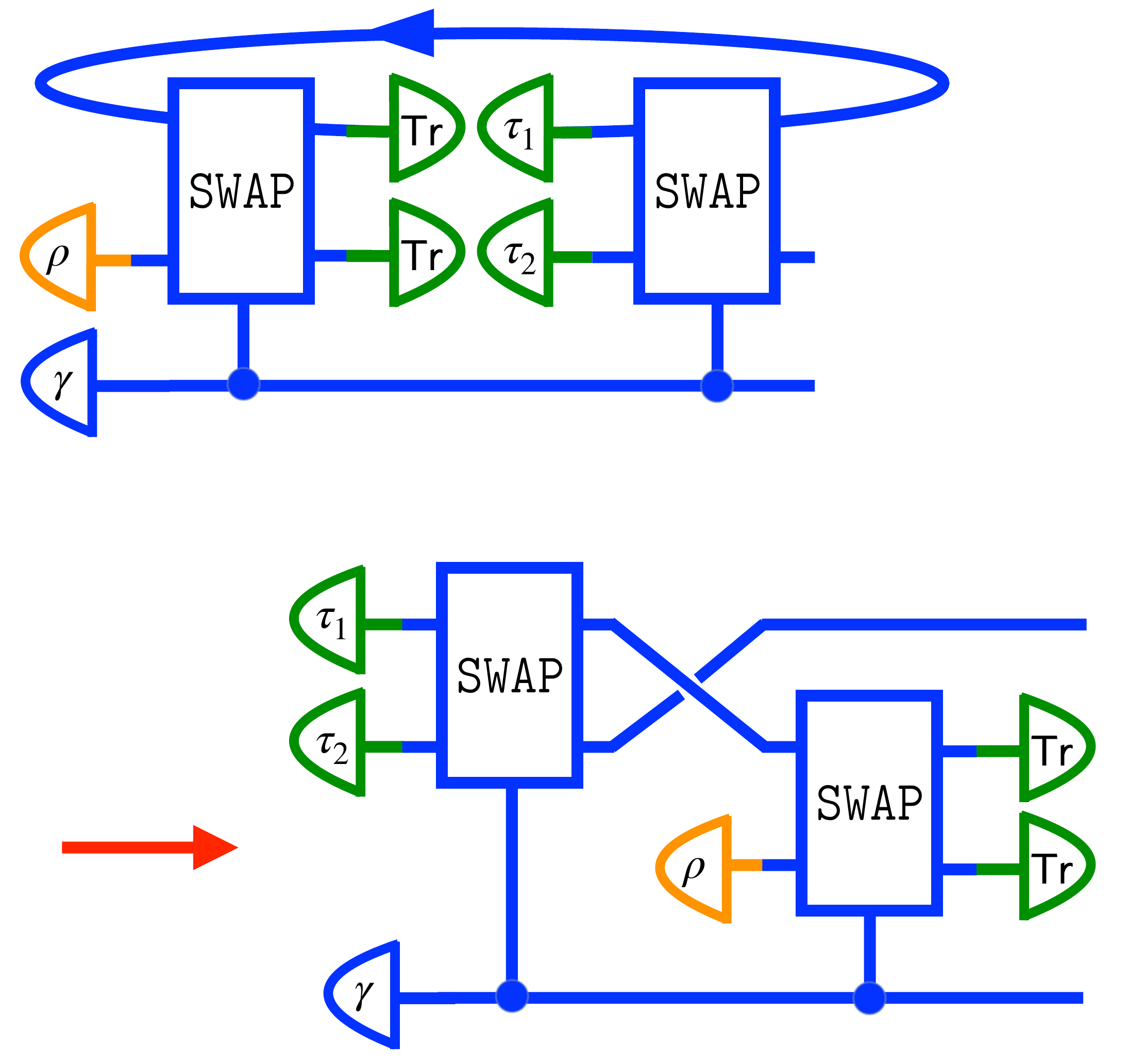} }
	\end{center}
	\caption{Unfolded quantum switch of constant channels.  The output of the quantum switch of two constant channels   (top)       can be reproduced  by a quantum circuit where the state of the target system undergoes controlled ${\tt SWAP}$ operations with two fixed states (bottom).   The two constant channels (in green) output fixed states $\tau_1$ and $\tau_2$, independently of their input.  The target system (in red) is initially in the state $\rho$, while the control qubit (in blue) is initially in the state $\gamma$.   }		\label{fig:unfolded}
\end{figure}
To gain an insight into the physical origin of the anomalous heat flow, it is helpful to consider a hypothetical realization of the quantum switch in terms of a closed timelike curve \cite{chiribella2009beyond}.  In this realization, the quantum switch of two channels is realized by putting the two channels in parallel, and by sandwiching them between two controlled {\tt SWAP} operations, corresponding to the unitary gate  $c-{\tt SWAP}=  I\otimes I\otimes |0\rangle\langle 0|  +  {\tt SWAP} \otimes |1\rangle\langle 1|$, where $\tt SWAP$ is the unitary operator defined by the relation ${\tt SWAP} (  |\phi\rangle \otimes |\psi\rangle)   =  |\psi\rangle \otimes |\phi\rangle$, for every pair of states $|\phi\rangle$ and $|\psi\rangle$.  One of the outputs of the controlled ${\tt SWAP}$ operations is then fed inside a closed timelike curve, modelled as postselected quantum teleportation. This realization of the quantum switch is shown in the top part of Fig. \ref{fig:unfolded}, for the special case in which the input of the quantum switch are two constant channels ({\em i.e.}, channels that prepare fixed output states independently of their input). 

For constant channels, the realization of the quantum switch in the top part of Fig. \ref{fig:unfolded} can be unfolded into a quantum circuit using controlled $\tt SWAP$ operations in a definite causal order, as illustrated in the bottom part of Fig. \ref{fig:unfolded} (see Supplementary Material for the detailed derivation). 
Compared to the quantum switch, this circuit features a distinct structure that does not rely on ICO.
Instead of taking the constant channels themselves as input, it acts on the two output states produced by those channels. Despite these structural differences, the circuit provides valuable insight into the physical origin of the anomalous heat flow. As will be shown in the subsequent analysis, this origin can be attributed to the implementation of controlled $\tt SWAP$ operations with the control system prepared in a coherent superposition state.      

For simplicity, consider two qubits initially prepared in thermal states $\tau_1$ and $\tau_2$, respectively. A controlled $\mathtt{SWAP}$ operation is then applied, with the control qubit in the state $|+\rangle_c$. After the operation, the control qubit is measured in the basis $\{|+\rangle_c, |-\rangle_c\}$, and the second target qubit is discarded. The resulting state of the first target qubit is given by
\begin{align}
	\rho_\pm  =   \frac{  \tau_1+  \tau_2  \pm  \tau_1^2 \tau_2   \pm  \tau_2 \tau_1^2   }{  2  (  1  \pm   {\mathrm{Tr} }  [ \tau_1^2 \tau_2  ])}  \, ,
\end{align}      
where the subscript $\pm$ refers to the two possible outcomes of the measurement on the control qubit. 
Anomalous heat flow between the two target systems can be observed in several regimes. For instance, when the two target qubits start at the same temperature, the final state conditioned on outcome $|-\rangle_c$ reaches infinite temperature, while the state conditioned on outcome $|+\rangle_c$ attains a temperature lower than the initial one. This indicates that heat has flowed between two systems initially in thermal equilibrium.

This situation arises due to the controlled {\tt SWAP} operation with the control qubit in the $|+\rangle_{c}$, or more generally, in a state with coherence with respect to the computational basis used to control the {\tt SWAP} operation.  This combination of controlled {\tt SWAP} operation and coherence in the control qubit injects thermodynamical resources into the target qubits, and is at the origin of the anomalous heat flow. This point can be better observed by assigning a Hamiltonian to the control qubit, and assessing the resourcefulness of the state preparation and controlled {\tt SWAP} operation. Suppose that the Hamiltonian of the control qubit is diagonal in the computational basis $\{|0\rangle_{c}, |1\rangle_{c} \}$. In that case, the controlled {\tt SWAP} operation is Gibbs-preserving: it does not change the thermal state of the two target qubits and of the control. Hence, no anomalous heat flows between the two target qubits takes place if the control is in the thermal state. In other words, anomalous heat flow between the target qubits is only possible if thermodynamic resources are injected from the control qubit. Alternatively, suppose that the Hamiltonian of the control qubit is not diagonal in the computational basis. In this case, the controlled {\tt SWAP} operation is not Gibbs-preserving and therefore its implementation must require the injection of thermodynamic resources.  In summary, anomalous heat flow between the target qubits requires thermodynamical resources to be injected in the realization of the controlled {\tt SWAP} operation, and/or in the preparation of the initial state of the control qubit.

The ICO scheme based on the quantum switch has a similar origin, as the quantum switch can be thought as a controlled {\tt SWAP} operation between two different time slots.  As shown in the previous discussions, controlled ${\tt SWAP}$ operations with the control in a coherent superposition state can generally drive anomalous heat flows. A similar effect is achieved by the quantum switch by coherently controlling the time slots in which two thermalization channels occur. In summary, the coherently controlled $\tt SWAP$ operation, whether governing the spatial arrangement of two states or the causal order of two channels, can induce anomalous heat flow. This effect stems from thermodynamic resources either present in the control system or implicitly required to implement the controlled operations.

It is noteworthy, however, that the anomalous heat flow induced by ICO
in this study cannot be reproduced simply by coherently controlling the
choice of thermalization process. Specifically, when the control qubit is
in state $\ket{0}_c$ ($\ket{1}_c$), the system interacts solely with channel 
$\mathcal{N}^{T_{\texttt{E}}}_1$ ($\mathcal{N}^{T_{\texttt{E}}}_2$). 
Even when the control qubit is prepared in the $\ket{+}_c$ state and subsequently measured in the $\{\ket{+}_c, \ket{-}_c\}$ basis, no anomalous heat flow emerges in this configuration—in contrast to what occurs under coherently controlled SWAP operations. A detailed derivation is provided in the Supplemental Material.

\begin{figure}[t]
	\begin{center}
{\includegraphics[width=\linewidth]{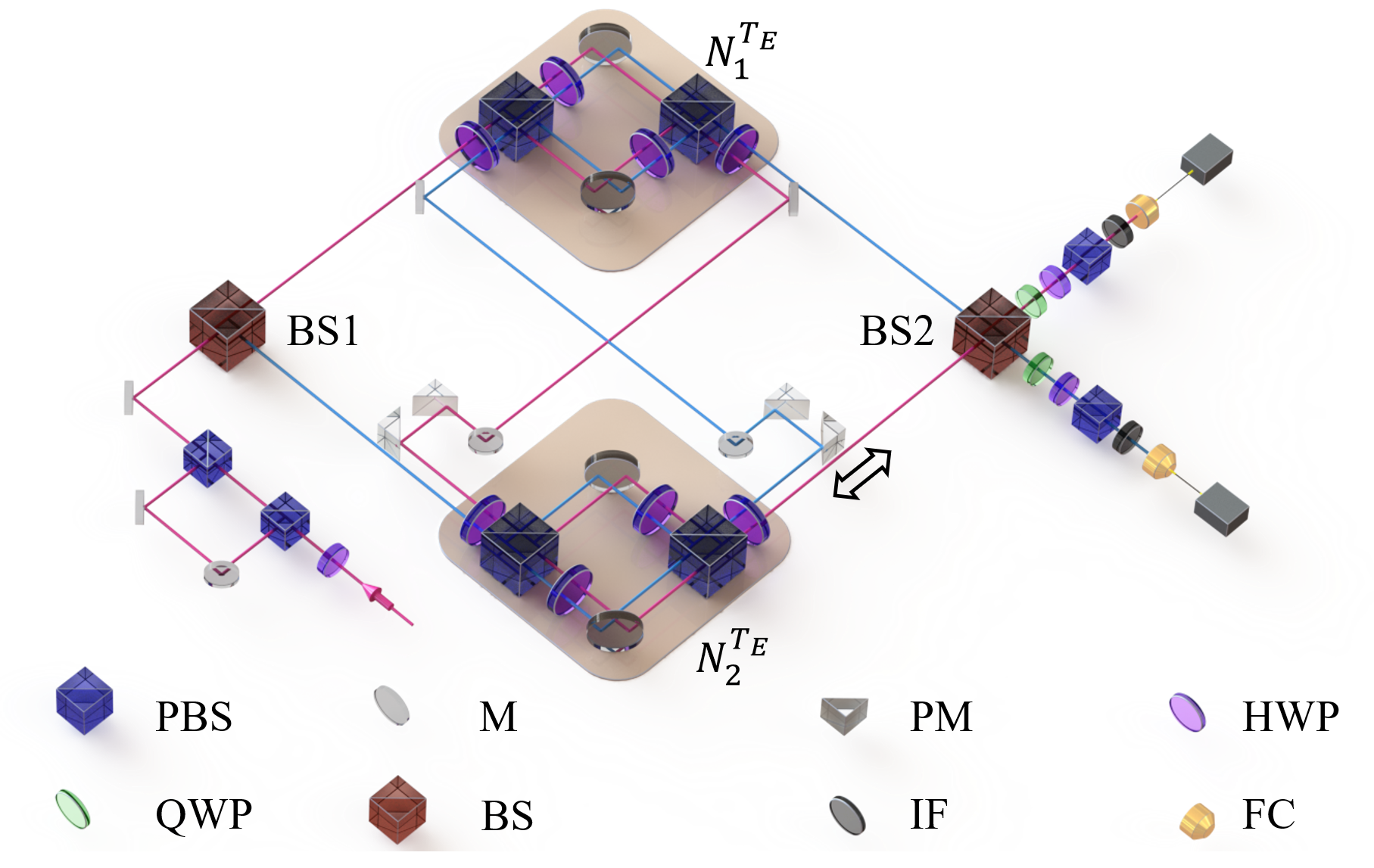} }
	\end{center}
	\caption{Experimental setup. The quantum switch is realized by
			a Mach-Zehnder interferometer structure, which comprises two
			equivalent thermalizing channels $\mathcal{N}^{T_{\texttt{E}}}_{1}$
			and $\mathcal{N}^{T_{\texttt{E}}}_{2}$ in the ICO.
			The two causal orders are characterized by red and blue optical paths, respectively.
			Polarization beam splitter (PBS), mirror (M), prism mirror (PM), 
			half wave plate (HWP), quarter-wave plate (QWP), beam splitter (BS), 
			interference filter (IF), fiber collimator (FC).}
	\label{Exp}
\end{figure}

\emph{\bfseries Experimental simulation of anomalous heat flow.}---
We employ a photonic quantum switch, as illustrated in Fig. \ref{Exp}, to simulate the ICO. The system and control qubits are encoded using the polarization and path degrees of freedom of photons, respectively. The ground state $\left|g\right\rangle$ and excited state $\left|e\right\rangle$ of the system correspond to the vertical ($\left|V\right\rangle$) and horizontal ($\left|H\right\rangle$) polarization states, respectively. The control qubit states $\left|0\right\rangle_c$ and $\left|1\right\rangle_c$ are represented by the blue and red optical paths.
The first beam splitter (BS1) introduces two spatial modes for the photons. In one interferometer arm (red path in Fig. \ref{Exp}), photons undergo the channel sequence $\mathcal{N}^{T_{\texttt{E}}}_{2} \circ
\mathcal{N}^{T_{\texttt{E}}}_{1}$, while in the other arm (blue path), they experience the reversed order $\mathcal{N}^{T_{\texttt{E}}}_{1} \circ \mathcal{N}^{T_{\texttt{E}}}_{2}$.
The second beam splitter (BS2) coherently superposes the two spatial modes and projects the control qubit onto the basis $\{\left|+\right\rangle_c, \left|-\right\rangle_c\}$.
At each output port of BS2, polarization analyzers are placed to perform quantum state tomography.
Further details on the optical simulation of the thermalizing channel are provided in the Supplemental Material. We conducted quantum process tomography on five distinct thermalizing channels at different temperatures, achieving an average process fidelity exceeding $99\%$, which confirms the reliability of our optical simulation of the thermalization channel.

\begin{figure}[t]
	\begin{center}
		\begin{tabular}{cccc}
		{\includegraphics[width=0.55\linewidth]{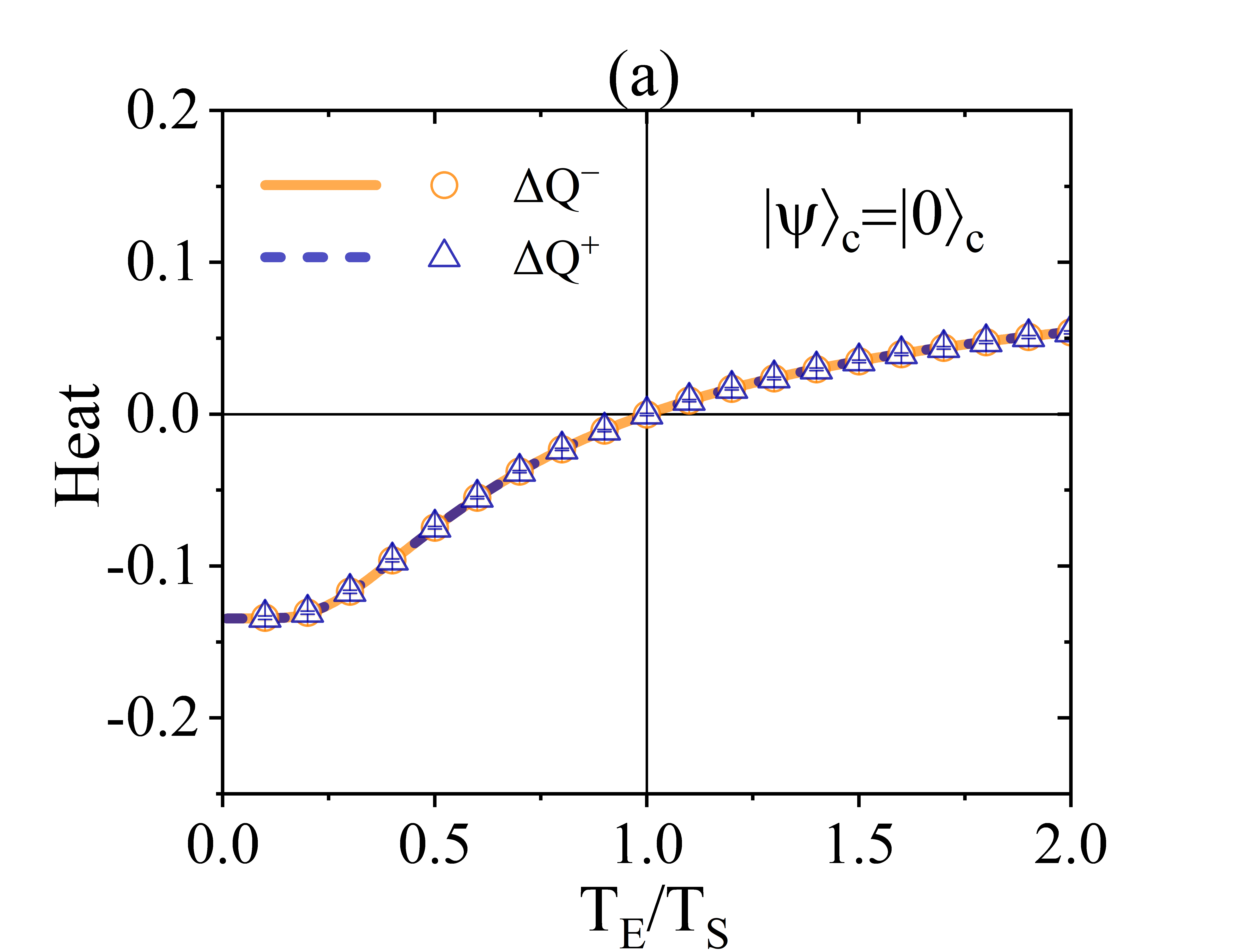} }&\hskip -0.6cm
		{\includegraphics[width=0.55\linewidth]{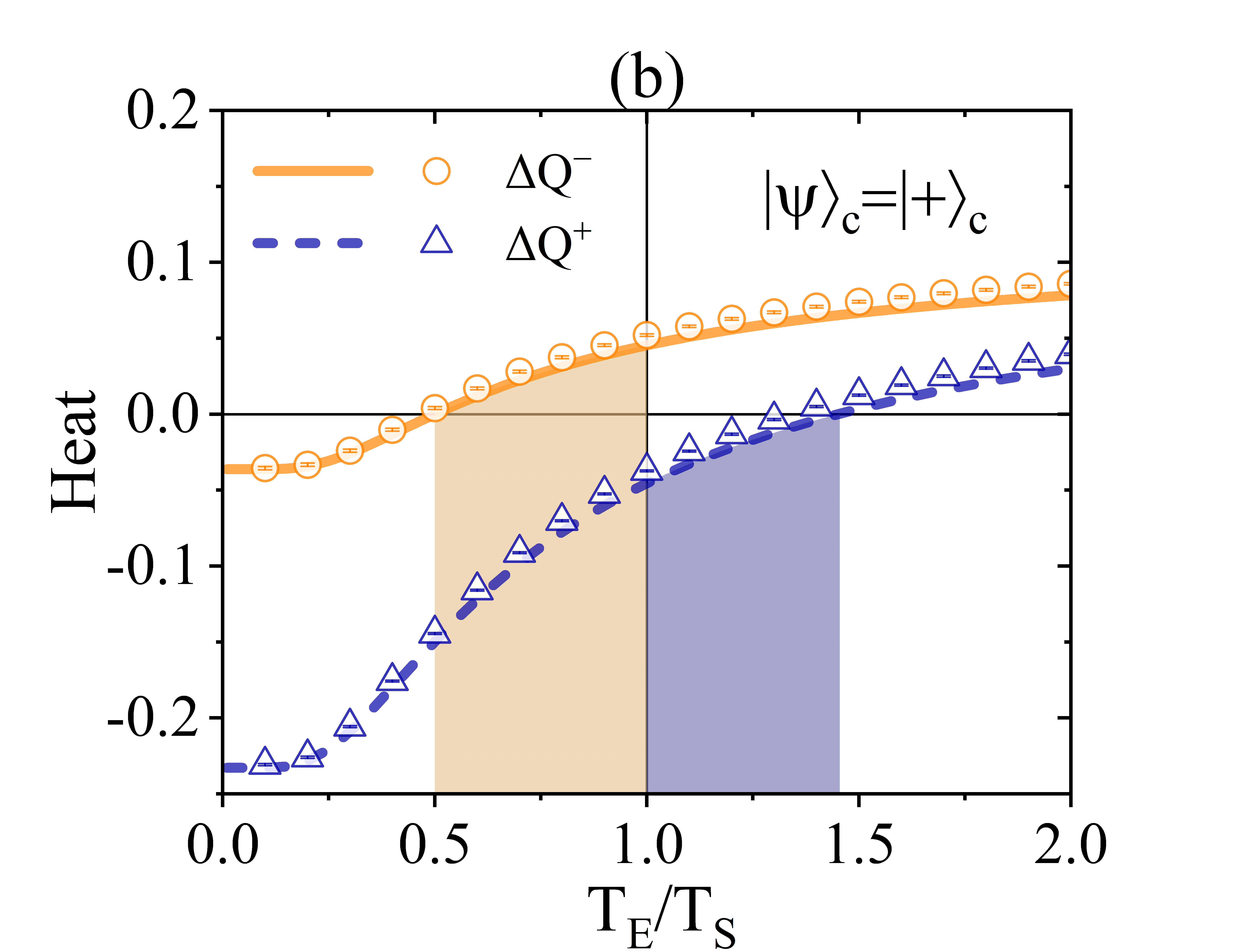} }&\\
		{\includegraphics[width=0.55\linewidth]{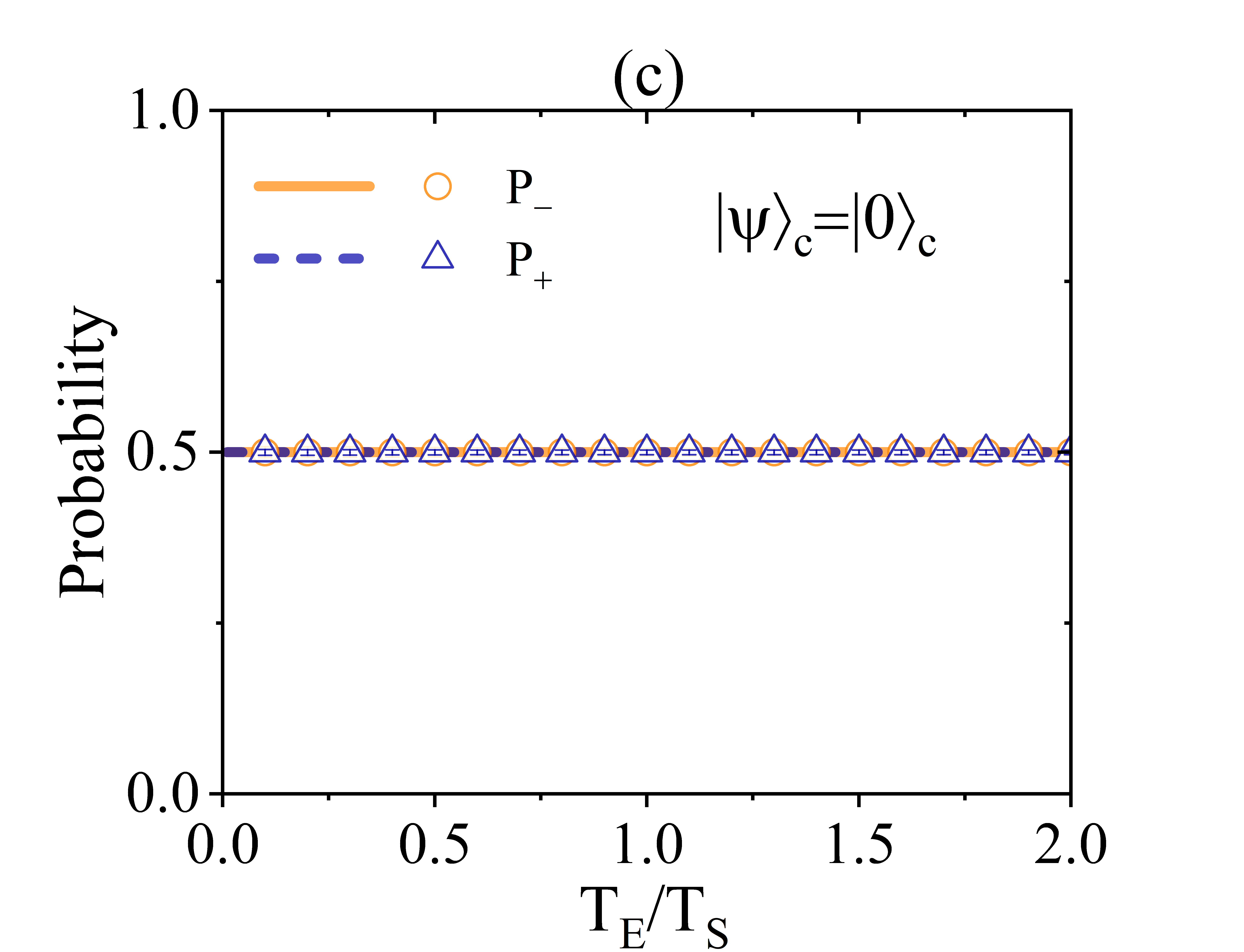} }&\hskip -0.6cm
		{\includegraphics[width=0.55\linewidth]{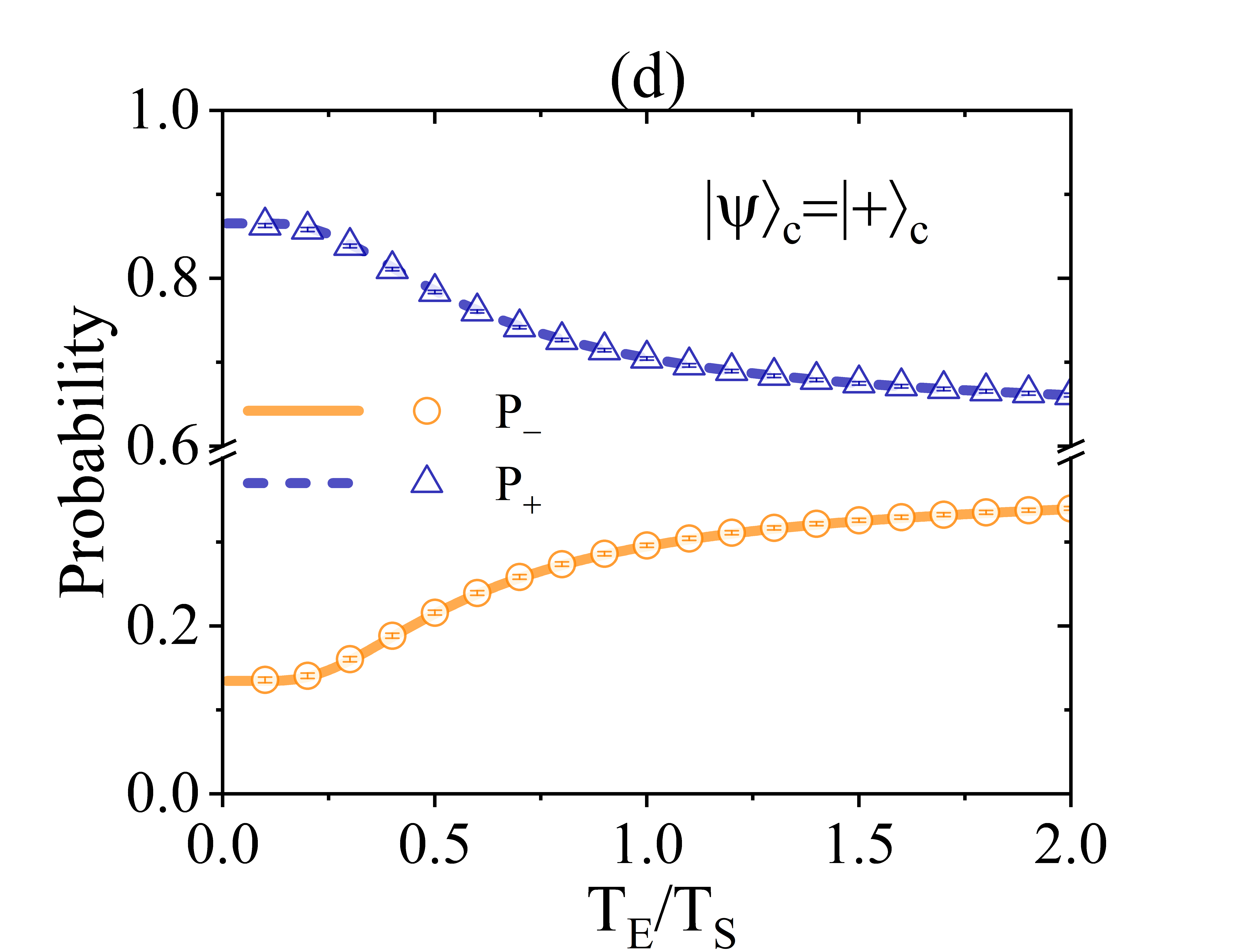} }&
		\end{tabular}
	\end{center}
	\caption{(a) and (b) Heat changes in the system, i.e., $\Delta Q^{\pm}$,
		against $T_{\texttt{E}}/T_{\texttt{S}}$, when the control qubit is measured in the
		$\{|+\rangle_c,|-\rangle_c\}$ basis with its 
		initial state set to $\ket{0}_{c}$ in (a) and $\ket{+}_{c}$ in (b), respectively.
		(c) and (d) The probabilities $P_{\pm}$ of measuring the control qubit
		in the $\ket \pm_{c}$ basis corresponding
		to scenarios (a) and (b), respectively. In all the figures, the curves
		denote theoretical predictions and the symbols represent the experimental data.
		}
	\label{heat}
\end{figure}

In the experiment, we illustrate the energy changes of the system
by varying the temperature $T_{\texttt{E}}$ of the channels.
As a reference scenario, we first consider a classical setup where the control qubit's initial state is fixed at $\left|0\right\rangle_{c}$.
Under this condition, an ICO cannot be established.
The system eventually arrives at a thermal state with temperature $T_{\texttt{E}}$,
regardless of the measurement outcome of the control qubit.
As a result, when $T_{\texttt{E}}<T_{\texttt{S}}$ $(T_{\texttt{E}}>T_{\texttt{S}})$, 
the system releases (absorbs) heat, denoted as $\Delta Q^{\pm}<0$
$(\Delta Q^{\pm}>0)$, to (from) the channels. This behavior is visually depicted in Fig. \ref{heat}(a).
Furthermore, the measurement probabilities $P_+$ and $P_{-}$,
which correspond to the control qubit being in the states
$\left|+\right\rangle_{c}$ and $\left|-\right\rangle_{c}$,
respectively, remain uniformly at $0.5$ throughout the process [see Fig. \ref{heat}(c)].

We then examine the ICO process
by choosing the initial state of the control qubit to $\left|+\right\rangle_{c}$.
After the system passing through the channels, the control qubit is measured
in the basis $\{|+\rangle_c,|-\rangle_c\}$. 
As illustrated in Fig. \ref{heat}(b), we observe a remarkably counterintuitive phenomenon: when  
$T_{\texttt{E}}/T_{\texttt{S}}<1$, heat can be transferred from the low-temperature
channels to the high-temperature system, resulting in $\Delta Q^{-}>0$,
provided that the control qubit is measured to be in the state $|-\rangle_c$ 
(as indicated by the cyan shaded area in the figure). Conversely, when $T_{\texttt{E}}/T_{\texttt{S}}>1$,
heat can flow from the low-temperature system to the hot-temperature
channels, leading to $\Delta Q^{-}<0$, given that the control qubit is measured
to be in the state $|+\rangle_c$
(as depicted by the blue shaded area in the figure).
The intervals of $T_{\texttt{E}}$
during which these anomalous heat flows occur are consistent with
the conditions (\ref{con1}) and (\ref{con2}) we derived: for $T_{\texttt{E}}/T_{\texttt{S}}<1$, 
the interval is $T_{\texttt{E}}/T_{\texttt{S}}>0.5$; and for  
$T_{\texttt{E}}/T_{\texttt{S}}>1$, it is approximately $T_{\texttt{E}}/T_{\texttt{S}}\lesssim1.45$.
The corresponding measurement probabilities $P_{\pm}$
of the control qubit in the basis $\{|+\rangle_c,|-\rangle_c\}$
are presented in Fig. \ref{heat}(d).

Although the ICO process can produce counterintuitive
heat flows, this does not violate the second law of thermodynamics. 
The occurrence of anomalous heat flow is strictly contingent on
the measurement of the control qubit. If the measurement outcomes are not recorded, 
the total heat exchange in the ICO process, denoted by
$(\Delta Q^{+}+ \Delta Q^{-})_\text{ICO}$, equals that of the classical process,
$(\Delta Q^{+}+\Delta Q^{-})_\text{clas}$. 
This is exemplified in the case where $T_{\texttt{E}}/T_{\texttt{S}}=1$:
the net heat exchange between the system and the channels remains zero in both scenarios,
i.e., $(\Delta Q^{+}+\Delta Q^{-})_\text{ICO}=(\Delta Q^{+}+\Delta Q^{-})_\text{clas}=0$.

We further observe that the presence of ICO modifies the magnitude of heat flow, even in regimes where its direction is not reversed, specifically, in the regions where $T_{\texttt{E}}/T_{\texttt{S}}<0.5$ and $T_{\texttt{E}}/T_{\texttt{S}}\gtrsim1.45$
in Fig. \ref{heat}(b). This stands in contrast to the case without ICO, shown in Fig. \ref{heat}(a). 
The heat flow depends on the probabilities $P^{\pm}$ which in turn are subjected to causal inequalities \cite{Araujo2015, Hoban2015} testifying the nature of ICO.
Moreover, it depends critically on the measurement outcome of the control qubit. For instance, in the region
$T_{\texttt{E}}/T_{\texttt{S}}<0.5$, 
we find $|\Delta Q^{+}|>|\Delta Q^{-}|$,
indicating that the system releases more heat to the channels when the control qubit is projected onto
$|+\rangle_{c}$. Conversely, for $T_{\texttt{E}}/T_{\texttt{S}}\gtrsim1.45$,
$\Delta Q^{-}>\Delta Q^{+}$ 
implies that the system absorbs more heat from the channels upon measuring $|-\rangle_{c}$.

\begin{figure}[b]
	\begin{center}
		{\includegraphics[width=\linewidth]{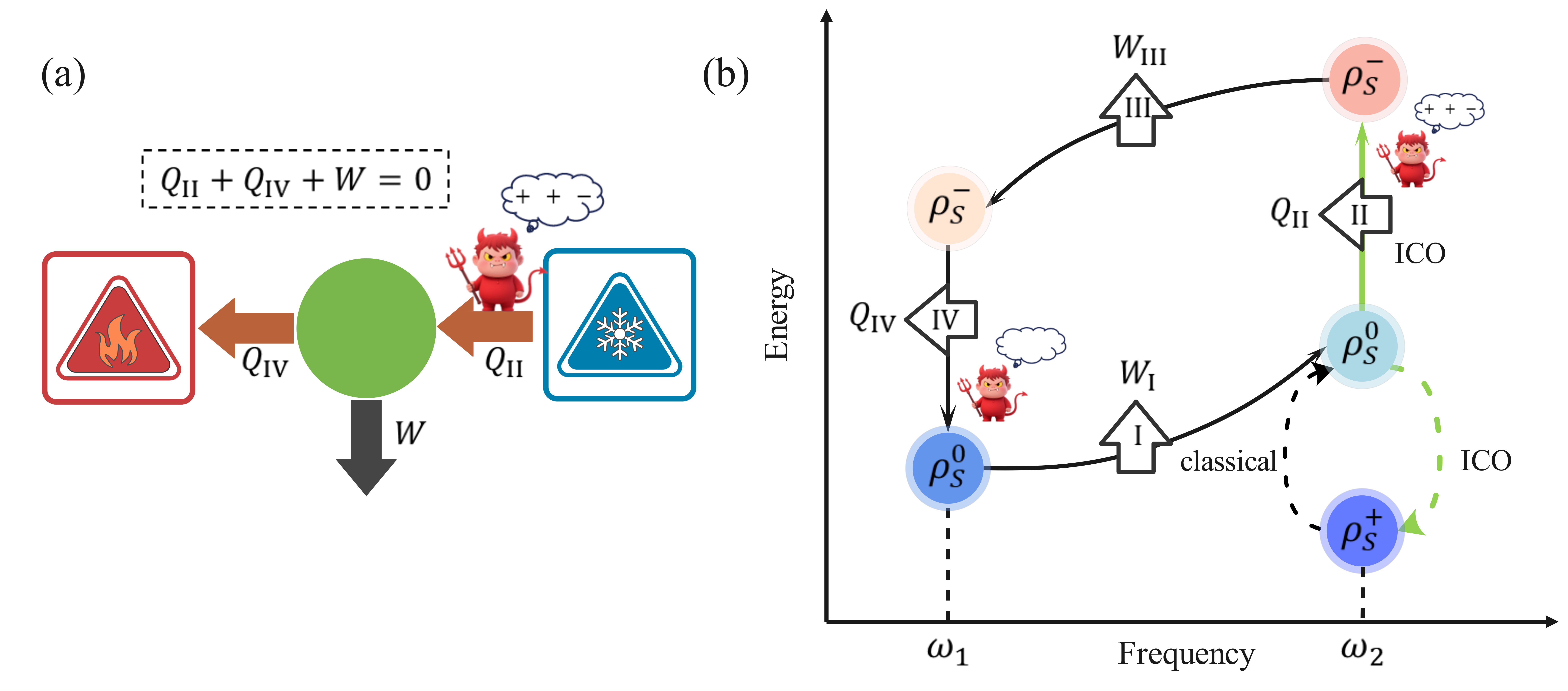} }
	\end{center}
	\vskip -0.3cm
	\caption{(a) Schematic of work-heat conversion in an ICO-based thermal machine. The work cost for erasing the memory of Maxwell's demon allows the system to absorb heat from a cold source. Part of this heat is transferred to a hot source, and the remainder is converted into work on an external agent.
(b) Illustration of the ICO-based Otto cycle. Strokes I (work input), III (work output), 
and IV (system cooling, combined with demon memory reset) are analogous to a standard Otto cycle.
The cycle diverges in Stroke II, wherein the system undergoes an ICO process (green, "ICO"),
followed by a measurement of the control qubit by the demon. A $|-\rangle_c$ result allows the cycle to continue; otherwise, the system is thermally reset via an auxiliary source (black dashed, ``classical") to the state after Stroke I.
This process iterates until a $|-\rangle_c$ measurement occurs.  }
	\label{OttoModel}
\end{figure}

\emph{\bfseries Quantum Otto cycle with ICO.}---
In this section, we examine the application potential of this anomalous heat flow. 
To this end, we address the question: can a machine be constructed that utilizes ICO to induce heat transfer from lower to higher temperatures while maintaining a net work output? We propose a machine design achieving this, illustrated in Fig. \ref{OttoModel}(a). Much like Maxwell's demon, our machine is powered by the work required to erase the demon's memory. 
It plays an analogous role of the plug in a classical fridge.

Specifically, we design a modified Otto cycle [Fig. \ref{OttoModel}(b)] that functions simultaneously as both a refrigerator and an engine. The working substance is a qubit described by the Hamiltonian
$H(t) = \omega(t) \sigma_{z}/2$, where $\omega(t)$ is a time-dependent transition frequency. 
A key distinction from the conventional Otto cycle lies in the heat exchange mechanism:
the system absorbs heat from the low-temperature channel $R_{\mathrm{\Rmnum{2}}}$ (at temperature $T_2$) via ICO, while releasing heat into the high-temperature channel $R_{\mathrm{\Rmnum{4}}}$ (at temperature $T_4$). 
The detailed cycle process can be depicted as follows.

\textbf{Stroke $\mathrm{\Rmnum{1}}$}: \textit{Quantum adiabatic compression}.
The system starts as a thermal state 
$\rho_{\texttt{S}}^{0}=f_{1}\left|e\right\rangle\left\langle e\right|
+(1-f_{1})\left|g\right\rangle\left\langle g\right|$ of temperature $T_{4}$
with $f_{1}=(e^{\omega_{1}/T_{4}}+1)^{-1}$.
An adiabatic transformation bring the frequency of the system from $\omega_1$ to $\omega_2 (> \omega_1)$ with an internal energy shift due to the external work
$W_{\mathrm{\Rmnum{1}}}=
\mathrm{Tr}\left[\rho_{\texttt{S}}^{0}\left(H_{2}-H_{1}\right)\right]$.
The temperature of the system is turned correspondingly from
$T_{4}$ to $T_{1}=T_{4}\omega_{2}/\omega_{1}$.

\textbf{Stroke $\mathrm{\Rmnum{2}}$}: \textit{Quantum isochoric thermalization driven by the ICO}. 
The system absorbs heat from the channel $R_{\mathrm{\Rmnum{2}}}$. 
In contrast to the conventional scenario, we set 
the temperature of $R_{\mathrm{\Rmnum{2}}}$  
to be lower than that of the system and even lower than $T_{4}$ of the
channel $R_{\mathrm{\Rmnum{4}}}$ in the fourth stoke,
namely, $T_{2}\leq T_{4}\leq T_{1}$.
We employ two equivalent channels $R_{\mathrm{\Rmnum{2}}}$
to interact with the system in ICO and prepare the control qubit in the 
state $|+\rangle_{c}$.
After the interactions, the control qubit is measured
in the basis $\{|+\rangle_{c},|-\rangle_{c}\}$.
Upon detecting the state $|-\rangle_{c}$, 
the system is
successfully heated, and a Maxwell's demon allows the cycle to continue.
Otherwise, we introduce a classical process where
the system is brought into contact with a thermalizing channel at temperature $T_{1}$,
returning the system to the state just after the stroke I (indicated by dashed lines in the diagram).
This step is repeated until the state $|-\rangle_{c}$ of the control qubit
is acquired.
The system then reaches the state $\rho_{\texttt{S}}^{-}=f^{-}\left|e\right\rangle\left\langle e\right|+(1-f^{-})\left|g\right\rangle\left\langle g\right|$,
in which $f^{-}=\frac{1}{2P^{-}}[f_{2}-2\sqrt{\alpha(1-\alpha)}f_{1}f_{2}^{2}]$ with
$f_{2}=(e^{\omega_{2}/T_{2}}+1)^{-1}$.
In this process, the heat absorbed by the system is 
$Q_{\mathrm{\Rmnum{2}}}=
\mathrm{Tr}\left[H_{2}\left(\rho_{\texttt{S}}^{-}-\rho_{\texttt{S}}^{0}\right)\right]$. 

\textbf{Stroke $\mathrm{\Rmnum{3}}$}: \textit{Quantum adiabatic expansion}.
The Hamiltonian of the system
is changed adiabatically from $H_{2}$ back to $H_{1}$ leaving
the state unchanged. The work done by the system to an external
agent is $W_{\mathrm{\Rmnum{3}}}
=\mathrm{Tr}\left[\rho_{\texttt{S}}^{-}\left(H_{1}-H_{2}\right)\right]$ and there is no heat exchange.

\textbf{Stroke $\mathrm{\Rmnum{4}}$}: \textit{Quantum isochoric cooling and information erasure}.
The system, with fixed Hamiltonian $H_{1}$,
is thermalized to the initial state $\rho_{\texttt{S}}^{0}$ by bringing it 
to interact with the channel $R_{\mathrm{\Rmnum{4}}}$ at
temperature $T_{4}$. 
The heat released by the system is 
$Q_{\mathrm{\Rmnum{4}}}=\mathrm{Tr}\left[H_{1}\left(\rho_{\texttt{S}}^{0}-\rho_{\texttt{S}}^{-}\right)\right]$.
To enable the cycle to continue,
we need to erase the memory of Maxwell's demon about the measurement information of
the control qubit.

\begin{figure}[t]
	\centering
	\begin{tabular}{cc}
		\includegraphics[width=0.55\linewidth]{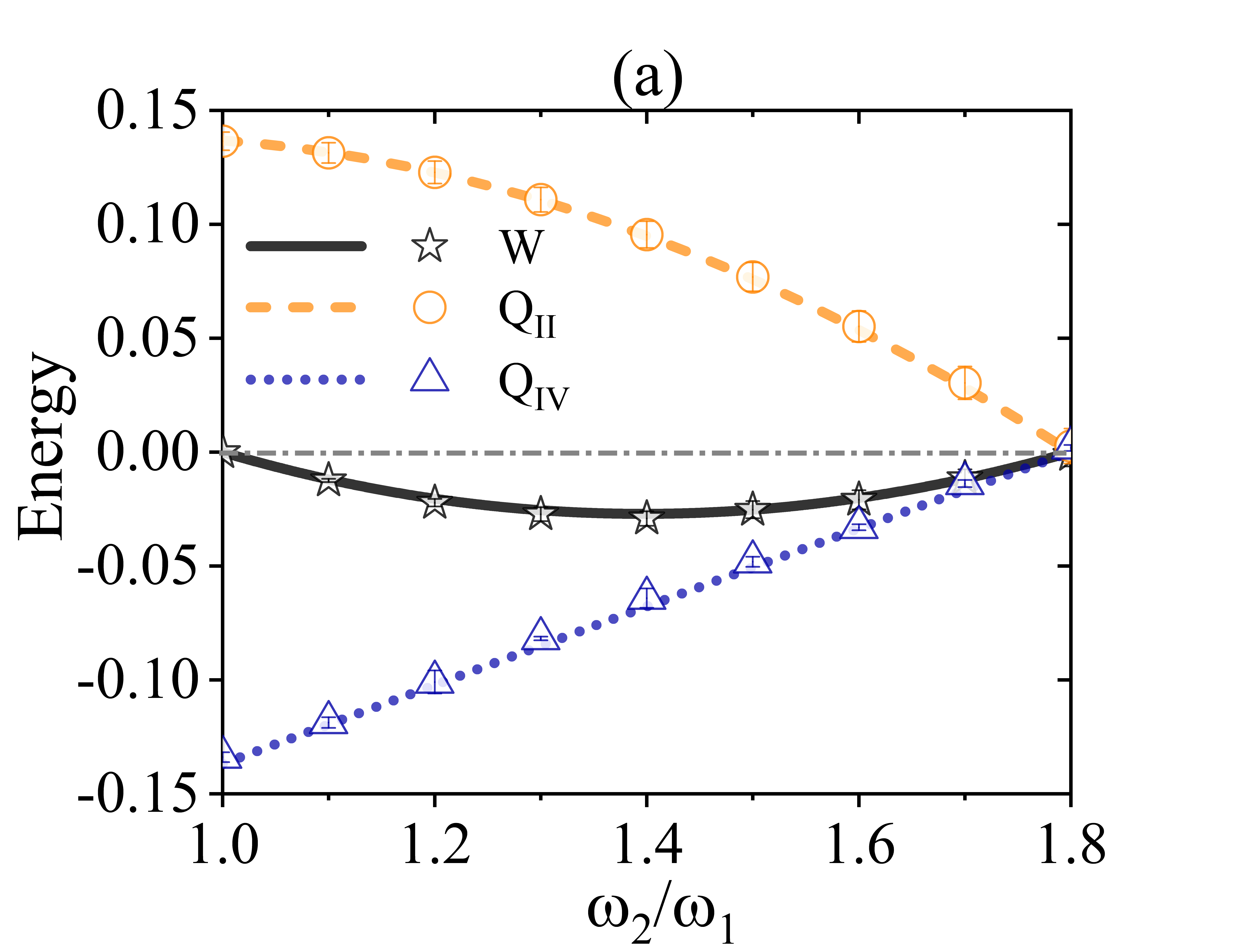} & \hskip -0.6cm
		\includegraphics[width=0.55\linewidth]{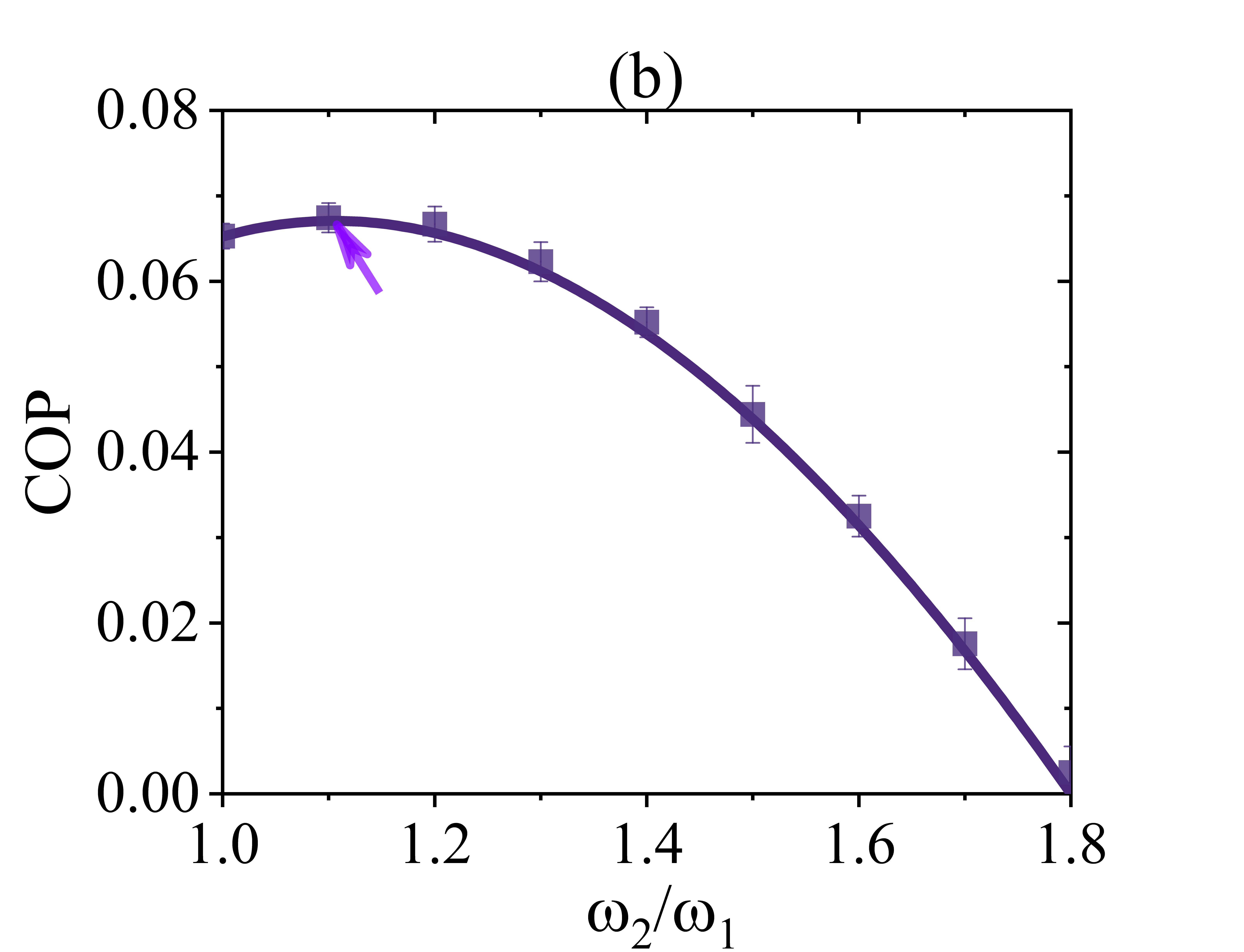} \\
	\end{tabular}
	\caption{(a) The net work $W$, the
		heat $Q_{\mathrm{\Rmnum{2}}}$ absorbed from the low-temperature channels $R_{\mathrm{\Rmnum{2}}}$,
		and the heat $Q_{\mathrm{\Rmnum{4}}}$ released to the high-temperature channel $R_{\mathrm{\Rmnum{4}}}$
		as a function of $\omega_{2}/\omega_{1}$. The temperatures of $R_{\mathrm{\Rmnum{2}}}$ and $R_{\mathrm{\Rmnum{4}}}$
		are chosen as $T_{2}=0.9\omega_{1}$ and $T_{4}=\omega_{1}$, respectively. 
		(b) The associated COP of the machine versus $\omega_{2}/\omega_{1}$. Here, 
		the arrow indicates the point where the COP reaches its maximum value.
		In both (a) and (b), the curves denote theoretical predictions and the symbols represent
		the experimental data.
}
	\label{machine}
\end{figure}

Using the common convention that a positive heat flow (work) corresponds to energy flowing into the system, coherently with our definitions, our machine is characterised by 
$Q_{\mathrm{II}}>0$, $Q_{\mathrm{IV}}<0$, and $W_{\mathrm{I}}+W_{\mathrm{III}}<0$ represents the net work in a cycle.
In addition, erasing the demon's memory requires work consumption,
$W_{\text{era}}=T_{r} \Delta S$
with $\Delta S=-\left(P^{-}\ln P^{-}+P^{+}\ln P^{+}\right)$
the Shannon entropy and $T_{r}$ the temperature of the 
reset reservoir in contact with the memory.
The coefficient of performance (COP) of this ICO-based
machine can be defined as the ratio of all useful outputs
to all energy inputs as
\begin{equation}\label{COP}
\mathrm{COP}=\frac{Q_{\mathrm{\Rmnum{2}}}+|W|}{W_{\text{era}}/P^{-}},
\end{equation}
in which the probability $P^{-}$ is taken into account, as 
the machine depends on obtaining the measurement outcome
$|-\rangle_{c}$ for the control qubit.

We simulate the entire cyclic process   
by means of a photonic setup which consists of five modules
implementing initial state preparation and the four strokes, respectively.
The experimental details are provided in the Supplementary Material.
In Fig. \ref{machine}(a), we exhibit the variations of work and heat of the 
machine with respect to $\omega_2/\omega_1$.
The machine absorbs heat from $R_{\mathrm{II}}$
($Q_{\mathrm{II}}>0$) and 
releases heat to $R_{\mathrm{I}}$
($Q_{\mathrm{IV}}<0$), while simultaneously performing net work on an external agent
($W<0$). Strikingly, by harnessing ICO, the device functions as a combined refrigerator and engine, thereby achieving heat transfer from a cold source to a hot sink while producing useful work output.

The COP of the machine is plotted against $\omega_{2}/\omega_{1}$ in Fig. \ref{machine}(b). 
Although the cooling power $Q_{\mathrm{II}}$ is maximized when no work is performed 
($\omega_{2}/\omega_{1}=1$, $W=0$), the COP itself reaches a maximum at
$\omega_{2}/\omega_{1} \approx 1.105$. At this optimal point, the machine both outputs work and absorbs heat from the low-temperature source.

\emph{\bfseries Conclusion.}---
In conclusion, we have utilized ICO to achieve anomalous heat flow,
enabling heat transfer from a low-temperature object to a high-temperature object. 
We have designed an ICO-based Otto cycle, demonstrating that such anomalous heat flow can be harnessed to realize a quantum machine capable of simultaneous refrigeration and work output. Through unfolding the quantum switch, 
we delved into the origin of this anomalous heat flow, confirming that ICO allows the setup to
access the free energy of the control qubit. 
Using a photonic quantum switch, we have simulated the anomalous heat flow and the ICO-based Otto cycle, providing a proof-of-principle demonstration of the theory.
Our research advances the application of ICO in quantum thermodynamics and offers insights for
developing heat machines that are unattainable by classical means.

\begin{acknowledgements}
	
Z. X. M. and Y. J. X. acknowledges support from 
National Natural Science Foundation (China) under Grants No. 12575024 and No. 12274257,
Natural Science Foundation of Shandong Province (China) under Grants
No. ZR2023LLZ015 and No. ZR2024LLZ012.
G.C. acknowledges support from the
Chinese Ministry of Science and Technology (MOST)
through grant 2023ZD0300600, from the Hong Kong Research Grant Council through grant 17310725, and from the State Key Laboratory of Quantum Information Technologies and Materials.
R.L.F. acknowledges support by MUR (Ministero
dell’Universit\`{a} e della Ricerca) through the following projects: PNRR Project ICON-Q – Partenariato Esteso NQSTI – PE00000023 – Spoke 2 – CUP: J13C22000680006, PNRR Project QUANTIP – Partenariato Esteso NQSTI – PE00000023 – Spoke 9 – CUP: E63C22002180006. R.L.F. and E.R. also acknowledge support from the PNRR Project PRISM – Partenariato Esteso RESTART – PE00000001 – Spoke 4 – CUP: E13C22001870001. E.R. thanks Kyrylo Simonov for insightful discussions.

\end{acknowledgements}

%

\appendix

\section{Theoretical derivation}

\subsection{The dynamics of the system under the ICO process}
For simplicity, we restrict the analysis to the case
where the system is a qubit with Hamiltonian $H_\texttt{S}=\frac{\omega_\texttt{S}}{2} \sigma_z$,
initially in a thermal state at temperature $T_\texttt{S}$, $\rho_\texttt{S}^{0}= (1-f_1) \ket{g}\bra{g} + f_1 \ket{e}\bra{e} $, with $\ket{g}$ $(\ket{e})$ denoting the ground (excited) state of the system
and $f_{1}=1/\left(1+e^{\omega_{\texttt{S}}/T_{\texttt{S}}}\right)$ ($\hbar=k_B=1$ throughout the paper).
Now take a second qubit, called control and suppose that the channels are applied
in the order $\mathcal{N}^{T_{\texttt{E}}}_{1}\circ\mathcal{N}^{T_{\texttt{E}}}_{2}$
when the control qubit is in $\left|0\right\rangle_{c}$,
and $\mathcal{N}^{T_{\texttt{E}}}_{2}\circ\mathcal{N}^{T_{\texttt{E}}}_{1}$
if in $\left|1\right\rangle_{c}$.
The ICO evolution map takes the form
\begin{equation}\label{QS}
	\mathbb{W}_{ij} = 
	\left| 0 \right\rangle_c \left\langle 0 \right| \otimes K_1^i K_2^j  
	+ \left| 1 \right\rangle_c \left\langle 1 \right| \otimes K_2^j K_1^i,
\end{equation}
where $K_{1(2)}^{i(j)}$ represent the Kraus operators for
the channels $\mathcal{N}_{1(2)}^{T_{\texttt{E}}}$ with the superscript 
$i$ $(j)$ denoting the number of the operators.

The interaction between a qubit and a finite-temperature thermal reservoir can be characterized by a generalized amplitude damping channel with the Kraus operators given as
\begin{eqnarray}\label{eq3}
{{E}_{0}}&=&\sqrt{p}\left( \left| e \right\rangle \left\langle  e \right|+\sqrt{1-r}\left| g \right\rangle \left\langle  g \right| \right),\nonumber\\ 
{{E}_{1}}&=&\sqrt{p}\sqrt{r}\left| e \right\rangle \left\langle  g \right|, \nonumber\\ 
{{E}_{2}}&=&\sqrt{1-p}\left( \sqrt{1-r}\left| e \right\rangle \left\langle  e \right|+\left| g \right\rangle \left\langle  g \right| \right), \nonumber\\ 
{{E}_{3}}&=&\sqrt{1-p}\sqrt{r}\left| g \right\rangle \left\langle  e \right|.  
\end{eqnarray}
where the parameters $p$ and $r$
denote the reservoir temperature and 
the interaction time between the qubit and reservoir, respectively.
${{E}_{0}}$ and ${{E}_{1}}$ correspond to the excitation process, 
while ${{E}_{2}}$ and ${{E}_{3}}$ correspond to the relaxation process.
In our model, the system experiences thermlizing channels which can be realized by setting 
$r=1$. The kraus operators of the thermalizing channel can thus be formulated as
\begin{eqnarray}\label{eq4}
&& K_{1(2)}^{0}=\sqrt{p}\left| e \right\rangle \left\langle  e \right|, K_{1(2)}^{1}=\sqrt{p}\left| e \right\rangle \left\langle  g \right|,\nonumber \\ 
&& K_{1(2)}^{2}=\sqrt{1-p}\left| g \right\rangle \left\langle  g \right|, K_{1(2)}^{3}=\sqrt{1-p}\left| g \right\rangle \left\langle  e \right|. 
\end{eqnarray}
By means of the definition $\mathbb{W}_{ij}$ given in Eq. (\ref{QS}),
the associated operators of the quantum switch can be constructed as
\begin{eqnarray}\label{eq5}
	&& {\mathbb{W}_{00}}=\left| 0 \right\rangle {{\left\langle  0 \right|}_{c}}\otimes p\left| e \right\rangle \left\langle  e \right|+\left| 1 \right\rangle {{\left\langle  1 \right|}_{c}}\otimes p\left| e \right\rangle \left\langle  e \right|  \nonumber\\ 
	&& {\mathbb{W}_{01}}=\left| 0 \right\rangle {{\left\langle  0 \right|}_{c}}\otimes p\left| e \right\rangle \left\langle  g \right| \nonumber\\ 
	&& {\mathbb{W}_{03}}=\left| 1 \right\rangle {{\left\langle  1 \right|}_{c}}\otimes \sqrt{p\left( 1-p \right)}\left| g \right\rangle \left\langle  e \right| \nonumber\\ 
	&& {\mathbb{W}_{10}}=\left| 1 \right\rangle {{\left\langle  1 \right|}_{c}}\otimes p\left| e \right\rangle \left\langle  g \right| \nonumber\\ 
	&& {\mathbb{W}_{12}}=\left| 0 \right\rangle {{\left\langle  0 \right|}_{c}}\otimes \sqrt{p\left( 1-p \right)}\left| e \right\rangle \left\langle  g \right| \nonumber\\ 
	&& {\mathbb{W}_{13}}=\left| 0 \right\rangle {{\left\langle  0 \right|}_{c}}\otimes \sqrt{p\left( 1-p \right)}\left| e \right\rangle \left\langle  e \right|\nonumber\\
	&&+\left| 1 \right\rangle {{\left\langle  1 \right|}_{c}}\otimes \sqrt{p\left( 1-p \right)}\left| g \right\rangle \left\langle  g \right| \nonumber\\ 
	&& {\mathbb{W}_{21}}=\left| 1 \right\rangle {{\left\langle  1 \right|}_{c}}\otimes \sqrt{p\left( 1-p \right)}\left| e \right\rangle \left\langle  g \right| \nonumber\\ 
	&& {\mathbb{W}_{22}}=\left| 0 \right\rangle {{\left\langle  0 \right|}_{c}}\otimes \left( 1-p \right)\left| g \right\rangle \left\langle  g \right|+\left| 1 \right\rangle {{\left\langle  1 \right|}_{c}}\otimes \left( 1-p \right)\left| g \right\rangle \left\langle  g \right| \nonumber\\ 
	&& {\mathbb{W}_{23}}=\left| 0 \right\rangle {{\left\langle  0 \right|}_{c}}\otimes \left( 1-p \right)\left| g \right\rangle \left\langle  e \right| \nonumber\\ 
	&& {\mathbb{W}_{30}}=\left| 0 \right\rangle {{\left\langle  0 \right|}_{c}}\otimes \sqrt{p\left( 1-p \right)}\left| g \right\rangle \left\langle  e \right| \nonumber\\ 
	&& {\mathbb{W}_{31}}=\left| 0 \right\rangle {{\left\langle  0 \right|}_{c}}\otimes \sqrt{p\left( 1-p \right)}\left| g \right\rangle \left\langle  g \right|\nonumber\\
	&&+\left| 1 \right\rangle {{\left\langle  1 \right|}_{c}}\otimes \sqrt{p\left( 1-p \right)}\left| e \right\rangle \left\langle  e \right| \nonumber\\ 
	&& {\mathbb{W}_{32}}=\left| 1 \right\rangle {{\left\langle  1 \right|}_{c}}\otimes \left( 1-p \right)\left| g \right\rangle \left\langle  e \right| \nonumber\\ 
	&& {\mathbb{W}_{02}}={\mathbb{W}_{11}}={\mathbb{W}_{20}}={\mathbb{W}_{33}}=0. 
\end{eqnarray}

The initial state of the control qubit is prepared as ${{\rho }_{c}}={{\left| \psi  \right\rangle }_{c}}\left\langle  \psi  \right|$ with ${{\left| \psi  \right\rangle }_{c}}=\sqrt{\alpha }{{\left| 0 \right\rangle }_{c}}+\sqrt{1-\alpha }{{\left| 1 \right\rangle }_{c}}$. 
By applying the quantum switch to the system with initial state $\rho _{\mathrm{S}}^{0}$,
the total state of the system and control qubit evolves into
\begin{widetext}
\begin{eqnarray}\label{eq2}
	\rho _{cs}^{\mathrm{ICO}}&=&\sum\limits_{i,j}{{\mathbb{W}_{ij}}}\left( {{\rho }_{c}}\otimes \rho _{\mathrm{S}}^{0} \right)\mathbb{W}_{ij}^{\dagger } \nonumber\\
&=&p\alpha \left| 0e \right\rangle \left\langle  0e \right|+p\left( 1-\alpha  \right)\left| 1e \right\rangle \left\langle  1e \right| +\left( 1-p \right)\alpha \left| 0g \right\rangle \left\langle  0g \right|+\left( 1-p \right)\left( 1-\alpha  \right)\left| 1g \right\rangle \left\langle  1g \right| \nonumber\\ 
	&& +{{f}_{1}}{{p}^{2}}\sqrt{\alpha \left( 1-\alpha  \right)}\left( \left| 0e \right\rangle \left\langle  1e \right|+\left| 1e \right\rangle \left\langle  0e \right| \right)  +\left( 1-{{f}_{1}} \right){{\left( 1-p \right)}^{2}}\sqrt{\alpha \left( 1-\alpha  \right)}\left( \left| 0g \right\rangle \left\langle  1g \right|+\left| 1g \right\rangle \left\langle  0g \right| \right).  
\end{eqnarray}
\end{widetext}

By performing measurement on the control qubit in the basis 
$\left\{\left|+\right\rangle_{c},\left|-\right\rangle_{c}\right\}$, with 
$\left|\pm\right\rangle_{c}=\frac{1}{\sqrt{2}}\left(\left|0\right\rangle_{c}\pm\left|1\right\rangle_{c}\right)$,
the system is projected to the state
\begin{equation}\label{eq1}
	\rho _{\mathrm{S}}^{\pm }=\frac{_{c}\left\langle \pm \right|\rho _{cs}^{\mathrm{ICO}}{{\left| \pm  \right\rangle }_{c}}}{\text{Tr}{{[}_{c}}\langle \pm |\rho _{cs}^{\mathrm{ICO}}{{\left| \pm  \right\rangle }_{c}}]}={{f}^{\pm }}\left| e \right\rangle \langle e|+(1-{{f}^{\pm }})\left| g \right\rangle \langle g|,
\end{equation}
if $\left| \pm \right\rangle _{c}$ is obtained. 
In Eq. (\ref{eq1}), $f^{\pm}=\frac{1}{2P^{\pm}}\left[f_{2}\pm2\sqrt{\alpha (1-\alpha )}f_{1}f_{2}^{2}\right]$ with
${{P}^{\pm }}=\text{Tr}{{[}_{c}}\langle \pm |\rho _{cs}^{\mathrm{ICO}}{{\left| \pm  \right\rangle }_{c}}]=\frac{1}{2}\pm \sqrt{\alpha \left( 1-\alpha  \right)}$
$\times\left[ \left( 1-{{f}_{1}} \right){{\left( 1-f_2 \right)}^{2}}+{{f}_{1}}{{f_2}^{2}} \right]$ the corresponding measurement probabilities. 
Remarkably, the state $\rho_{\texttt{S}}^{\pm}$ of the system after interacting
with this ICO 
is related to its initial state through the parameter $f_{1}$.
This stands in sharp contrast to the typical thermalized state
$\rho_{\texttt{S}}^{T_{\texttt{E}}}=f_{2}
\left| e \right\rangle \left\langle e\right|+(1-f_2)\left| g \right\rangle \left\langle g\right|$
that the system would attain in the absence of ICO.
Moreover, the effective temperature of the states $\rho_{\texttt{S}}^{\pm}$
depends on the measurement outcomes of the control qubit.
This characteristic allows us to leverage ICO to manipulate the system's
temperature and accomplish a range of intriguing thermodynamic tasks.

The heat change of the system after the ICO process
is given as $\Delta Q^{\pm}=P_{\pm}\mathrm{Tr} \left[\left(\rho_{\texttt{S}}^{\pm}-\rho_{\texttt{S}}^{0}\right)H_{\texttt{S}}
\right]=P_{\pm}\omega_{\mathrm{S}}(f^{\pm}-f_{1})$. In contrast, if we set the initial state of the
control qubit as ${{\left| \psi  \right\rangle }_{c}}={{\left| 0 \right\rangle }_{c}}$,
the system will undergo the thermalizing channels $\mathcal{N}_{2}^{T_{\mathrm{E}}}$
and $\mathcal{N}_{1}^{T_{\mathrm{E}}}$ in sequence.  
By detecting the control qubit in the basis $\left\{\left|+\right\rangle_{c},\left|-\right\rangle_{c}\right\}$,
the system definitely collapses to the state
\begin{equation}
	\rho _{\mathrm{S}}^{\pm }=p\left| e \right\rangle \langle e|+(1-p)\left| g \right\rangle \langle g|
\end{equation}
with the same probability $1/2$ regardless of the measured results. 
In this case, 
the heat change of the system is given as
$\Delta Q^{\pm }=\frac{1}{2}\omega_{\mathrm{S}}\left(p-f_{1}\right)$.

\subsection{Derivation of the conditions for the occurrence of anomalous heat flow}
We first derive the condition under which the system is heated after undergoing
the ICO process when the initial temperature of the channels is lower than that of the system 
(i.e., $T_{\mathrm{E}}<T_{\mathrm{S}}$ or $f_2<f_1$). 
In the subsequent derivation, we take $\alpha=1/\sqrt{2}$ for the initial state of the 
control qubit. 
Since in this case the system's temperature after the ICO process
is higher when the control qubit is measured
in the state $\left|-\right\rangle_{c}$ than when it is measured in 
$\left|+\right\rangle_{c}$, i.e., $f^{-}>f^{+}$, the condition for the system to be heated
is reduced to $f^{-}>f_1$. 
By substituting $f^{-}=\frac{1}{2P^{-}}\left[f_{2}-
2\sqrt{\alpha (1-\alpha)}f_{1}f_{2}^{2}\right]$ into $f^{-}>f_1$,
we obtain $f_2>\frac{f_{1}^{2}}{1-2f_1+2f_{1}^{2}}$. Further substituting 
$f_1=1/\left({1+e^{\omega_{\mathrm{S}}/T_{\mathrm{S}}}}\right)$
and $f_2=1/\left({1+e^{\omega_{\mathrm{S}}/T_{\mathrm{E}}}}\right)$ 
into this inequality, 
we finally derive the condition under which the system can
be heated via ICO is $T_{\mathrm{S}}<2T_{\mathrm{E}}$.

Next, we derive the condition for the system to be cooled after undergoing
the ICO process when the initial temperature of the system is lower than the channels' temperature
(i.e., ${{T}_{\mathrm{S}}}<{{T}_{\mathrm{E}}}$ or ${{f}_{1}}<{{f}_{2}}$). Since the system's
temperature after ICO
is lower when the control qubit is measured in the state ${{\left|+\right\rangle}_{c}}$ than when it is measured in the state
${{\left|-\right\rangle}_{c}}$, i.e., ${{f}^{+}}<{{f}^{-}}$,
the condition for the system to be cooled is reduced to ${{f}^{+}}<f_1$. 
By substituting ${{f}^{+}}=\frac{1}{2{{P}^{+}}}[{{f}_{2}}+2\sqrt{\alpha\left(1-\alpha\right)}
{{f}_{1}}f_{2}^{2}]$ into ${{f}^{+}}<f_1$,
we obtain ${{f}_{2}}<1+\frac{1-f_{1}^{2}}{-1-2{{f}_{1}}+2f_{1}^{2}}$. 
By substituting the concrete forms of ${{f}_{1}}$
and $f_{2}$ into the inequality, we finally arrive at the condition for the system to be cooled
is $
T_{{\mathrm{E}}}<\omega_{\mathrm{S}}/\left( 2 \operatorname{artanh}
\left( \sinh\left( \frac{\omega_{\mathrm{S}}}{T_{\mathrm{S}}} \right)/
\left(\cosh\left(\frac{\omega_{\mathrm{S}}}{T_{\mathrm{S}}}\right)+2\right) \right)\right)$.
Taking $\omega_{\mathrm{S}}=1$ and $T_{\mathrm{S}}=1$,
we find that cooling of the system can be achieved when the channel temperature is less than 1.45043.

\subsection{The Otto cycle with channels of identical temperatures}

\begin{figure}[t!]
	\centering
	\begin{tabular}{c}
		\includegraphics[scale=0.25]{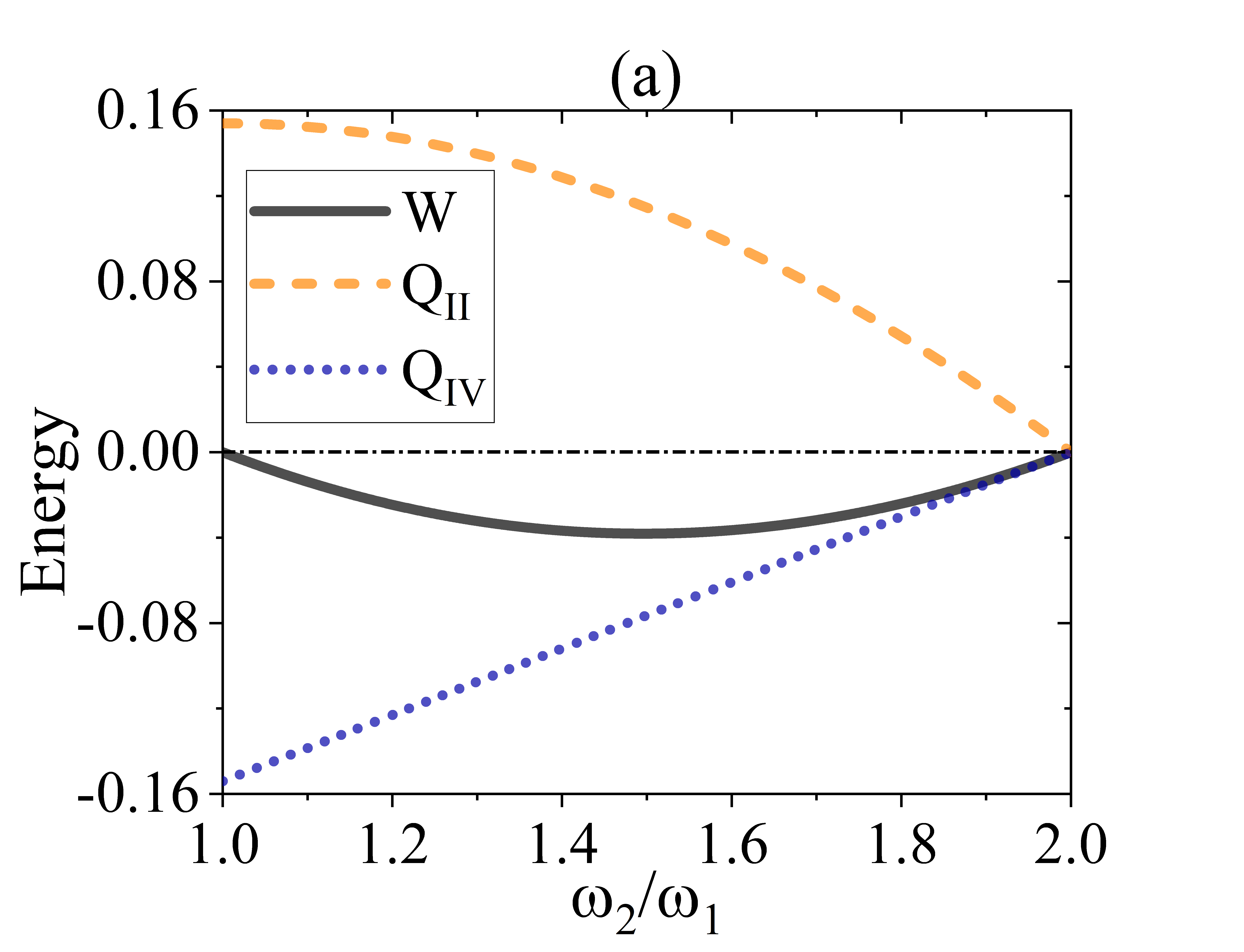}\\
		\includegraphics[scale=0.25]{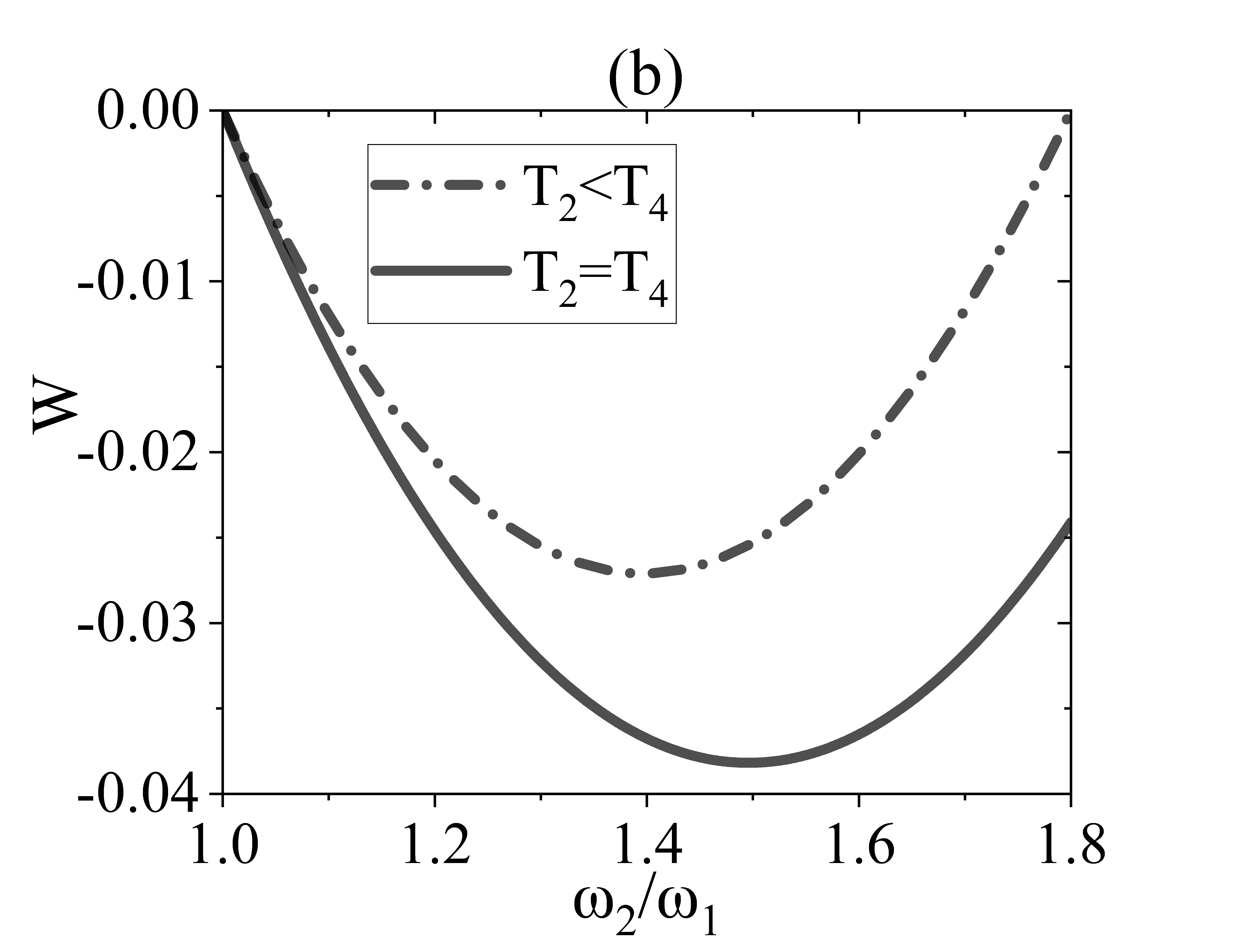}
	\end{tabular}
	\caption{(a) The net work $W$, the
		heat $Q_{\mathrm{\Rmnum{2}}}$ absorbed from the channels $R_{II}$ via ICO process,
		and the heat $Q_{\mathrm{\Rmnum{4}}}$ released to $R_{IV}$
		as a function of $\omega_{2}/\omega_{1}$. The temperatures of $R_{II}$ and $R_{IV}$
		are chosen as $T_{2}=T_{4}=\omega_{1}$.
		(b) A comparison for the new work $W$ for $T_2<T_4$ and $T_2=T_4$. For the case 
		$T_2<T_4$, we have chosen $T_{2}=0.9\omega_{1}$ and $T_{4}=\omega_{1}$. }\label{T2EqT4}
\end{figure}

In the main text, we have demonstrated the ICO-based Otto cycle, where
the heat can be transferred 
from the lower-temperature channels, $R_{II}$ at temperature $T_{2}$,
to the higher-temperature channel, $R_{IV}$
at temperature $T_{4}$ (with $T_{2}<T_{4}$), while simultaneously extracting work from the system.
In the following discussion, we relax the previous condition and consider the case 
of $T_2 = T_4$. In the cycle, the system's temperature starts from $T_{4}$,
which becomes $T_1=T_4 (\omega_{2}/\omega_{1})>T_4=T_{2}$
after undergoing the first adiabatic compression stroke.
Hence, in the second stroke, the system, with temperature $T_{1}$,
still needs to utilize the ICO process
to absorb heat from the channels $R_{II}$ at temperature $T_2<T_{1}$.

In Fig.\ref{T2EqT4} (a), we focus on the case where $T_{2}=T_{4}$ and
present the net work $W$, as well as the
heats $Q_{\mathrm{\Rmnum{2}}}$ and $Q_{\mathrm{\Rmnum{4}}}$,
as a function of $\omega_{2}/\omega_{1}$.
It can be observed that the system operates as an engine by absorbing heat from the channels
$R_{II}$ through the ICO process, with $Q_{\mathrm{\Rmnum{2}}}>0$.
Part of this absorbed heat is converted into useful work, indicated by  
$W<0$, while the remaining portion is released into 
channel $R_{IV}$ with $Q_{\mathrm{\Rmnum{4}}}<0$.
Compared with the case of 
$T_{2}<T_{4}$ studied in the main text, the system, as an engine,
can extract more net work under the case  
$T_{2}=T_{4}$, as shown in Fig. \ref{T2EqT4}(b).
This result aligns with our expectations,
since increasing the temperature of reservoir 
$R_{II}$ enhances the capacity of the system to produce useful work.
However, it should also be noted that when $T_{2}<T_{4}$,
the system is capable of not only performing work on the surroundings but also transferring heat
from the low-temperature place to the high-temperature one,
a feature that is not available under the present situation.

\section{The unfolded quantum switch}
Here we show that the quantum switch of two constant channels
can be simulated by a quantum circuit consisting of control-{\tt SWAP} operations
applied to the two states prepared by the two channels, as in Fig. 2 of the main text.  

Let $\map C_1$ and $\map C_2$ be the two constant channels, and let $\tau_1$ and $\tau_2$ be their output states, namely $\map C_1  (\rho)  =  \tau_1\, , \forall \rho$ and $\map C_2  (\rho)  =  \tau_2 \, \forall \rho$.  A Kraus representation of these two channels is provided by the operators   $C_{1mn}   =   \sqrt {\tau_1}  |m\>\<n| $ and $C_{2kl}   =   \sqrt {\tau_2}  |k\>\<l|$. 
The application of the quantum switch to the channels $\map C_1$ and $\map C_2$ yields a new channel $\map S (\map C_1,\map C_2)$ with Kraus operators 
\begin{align}
	S_{mnkl}    =  C_{1mn} C_{2 kl} \otimes |0\>\<0|   +  C_{2kl}  C_{1mn } \otimes |1\>\<1|  \, . 
\end{align}   
Explicitly, the action of the channel $\map S (\map C_1,\map C_2)$ on a product state $\rho \otimes \gamma$ is given by 
\begin{widetext}
\begin{eqnarray}\label{outputswitch}
	\map S (\map C_1,\map C_2)  &  = & c_{00} \,   \tau_1   \otimes |0\>\<0|     +  c_{11} \,   \tau_2   \otimes |1\>\<1| \nonumber \\ 
	& & +  c_{01}  \,  \sum_{m,n,k,l}  \sqrt{ \tau_1} |m\>\<n|   \sqrt{\tau_2}  |k\>\<l|  \rho  |n\>\<m|     \sqrt{ \tau_1}   |l\>\<k|  \sqrt{\tau_2}  \otimes |0\>\<1| \nonumber \\
	& & +  c_{10}  \,  \sum_{m,n,k,l}    \sqrt{\tau_2}  |k\>\<l|   \sqrt{ \tau_1} |m\>\<n|    \rho    |l\>\<k|  \sqrt{\tau_2}    |n\>\<m|     \sqrt{ \tau_1}  \otimes |1\>\<0| \nonumber \\ 
	& & =  c_{00} \,   \tau_1   \otimes |0\>\<0|     +  c_{11} \,   \tau_2   \otimes |1\>\<1|      +  c_{01}  \,   \tau_1  \rho   \tau_2  \otimes |0\>\<1|    +  c_{10}  \,  \tau_2   \rho  \tau_1 \otimes |1\>\<0| , 
\end{eqnarray}
\end{widetext}
where $(c_{ij})$ are the matrix elements of the density matrix $\gamma$.  

Now, we show that the action of the channel  $\map S (\map C_1,\map C_2)$ can be reproduced by a circuit using controlled-{\tt SWAP} operations, as in Fig. 2 of the main text.  
Let us denote by $E_1$ and $E_2$ the two environments, by $T$ the target system, and by $C$ the control qubit in the bottom part of Fig. 2 of the main text.  
We now follow the evolution of a pure product state  $|\psi_1\>_{E_1} \otimes |\psi_2\>_{E_2} \otimes |\phi\>_{T} \otimes |\gamma\>_{C}$ through the circuit at the bottom of Fig. 2 of the main text. Writing the state of the control qubit as $|\gamma\>  =  c_0  \, |0\>  +  c_1\,  |1\>$, the action of the first control swap in the circuit can be written as 
\begin{eqnarray}
	&{\tt cSWAP}_{E_1E_2C}  (|\psi_1\>_{E_1} \otimes |\psi_2\>_{E_2}\otimes |\gamma\>_{C}) &\nonumber\\
	 &  =  c_0  \,  |\psi_1\>_{E_1} \otimes |\psi_2\>_{E_2}\otimes |0\>_{C} &\nonumber\\
	 & + c_1  \,  |\psi_2\>_{E_1} \otimes |\psi_1\>_{E_2}\otimes |1\>_{C} \, . &
\end{eqnarray}
Then, the states of the two environments are swapped, obtaining the state 
\begin{eqnarray}
	&{\tt SWAP}_{E_1E_2}   {\tt cSWAP}_{E_1E_2C}  (|\psi_1\>_{E_1} \otimes |\psi_2\>_{E_2}\otimes |\gamma\>_{C})    & \nonumber\\
	&=  c_0  \,  |\psi_2\>_{E_1} \otimes |\psi_1\>_{E_2}\otimes |0\>_{C}  & \nonumber\\
	& + c_1  \,  |\psi_1\>_{E_1} \otimes |\psi_2\>_{E_2}\otimes |1\>_{C} \, . &  
\end{eqnarray} 
At this point, the second control swap acts on systems $E_2$, $T$ and $C$, yielding the state
\begin{widetext}
\begin{eqnarray}
&{ \tt cSWAP}_{E_2 TC}  {\tt SWAP}_{E_1E_2}    {\tt cSWAP}_{E_1E_2C}  (|\psi_1\>_{E_1} \otimes |\psi_2\>_{E_2}\otimes  |\phi\>_T  \otimes |\gamma\>_{C})  	\nonumber  \\
	&  \qquad  =  c_0  \,  |\psi_2\>_{E_1} \otimes |\psi_1\>_{E_2} \otimes |\phi\>_T  \otimes |0\>_{C}  + c_1  \,  |\psi_1\>_{E_1} \otimes |\phi\>_{E_2}  \otimes |\psi_2\>_{T}\otimes |1\>_{C} \, .   
\end{eqnarray}
\end{widetext}
Finally, we take the density matrix corresponding to the above pure state, and apply a partial trace on systems $E_2$ and $T$, obtaining the final state 
\begin{widetext}
\begin{eqnarray}
&\mathrm{Tr}_{E_2 T}  \left[ { \tt cSWAP}_{E_2 TC}  {\tt SWAP}_{E_1E_2}    {\tt cSWAP}_{E_1E_2C}  (|\psi_1\>\<\psi_1|_{E_1} \otimes |\psi_2\>\<\psi_2|_{E_2}\otimes  |\phi\>\<\phi|_T  \otimes |\gamma\>\<\gamma|_{C})    {\tt cSWAP}_{E_1E_2C}  {\tt SWAP}_{E_1E_2} { \tt cSWAP}_{E_2 TC}  \right]  &\nonumber \\  
&  \qquad  =   |c_0|^2  \,   |\psi_2\>\<\psi_2| \otimes |0\>\<0|   +   |c_1|^2  \,   |\psi_1\>\<\psi_1| \otimes |1\>\<1| 	\nonumber    \\
 &  \qquad  \quad +   c_0\overline c_1\,    \<\phi|\psi_1\>  \<  \psi_2|\phi\>  ~      |\psi_2\>\<\psi_1|  \otimes |0\>\<1|   	\nonumber   \\
&  \qquad   \quad +   c_1\overline c_0\,    \<\phi|\psi_2\>  \<  \psi_1|\phi\>  ~      |\psi_1\>\<\psi_2|  \otimes |1\>\<0| 	\nonumber  \\
 &  \qquad  =   |c_0|^2  \,   |\psi_2\>\<\psi_2| \otimes |0\>\<0|   +   |c_1|^2  \,   |\psi_1\>\<\psi_1| \otimes |1\>\<1| 	\nonumber    \\
	& \qquad +     c_0\overline c_1\,|\psi_2\>\<\psi_2| \phi\>\<\phi|\psi_1\>\<\psi_1|  \otimes |0\>\<1|+   c_1\overline c_0\,          |\psi_1\>  \<  \psi_1|\phi\>\<\phi|\psi_2\> \<\psi_2|  \otimes |1\>\<0|       \,. 
\end{eqnarray}
\end{widetext}
More generally, the output of the circuit for a product mixed state is 
\begin{widetext}
\begin{eqnarray}
	&\mathrm{Tr}_{E_2 T}  \left[ { \tt cSWAP}_{E_2 TC}  {\tt SWAP}_{E_1E_2}    {\tt cSWAP}_{E_1E_2C}  (\tau_{1} \otimes |\tau_{2}\otimes  \rho  \otimes \gamma)    {\tt cSWAP}_{E_1E_2C}  {\tt SWAP}_{E_1E_2} { \tt cSWAP}_{E_2 TC}  \right]  &\nonumber \\  
  &  =   c_{00}  \,   \tau_2 \otimes |0\>\<0|   +   c_{11}\,   \tau_1  \otimes |1\>\<1|  +     c_{01}  \,\tau_2 \rho \tau_1  \otimes |0\>\<1|+  c_{10} \,         \tau_1\rho \tau_2 \otimes |1\>\<0|, &
\end{eqnarray}
\end{widetext}
where $(c_{mn})$ are the entries of the density matrix $\gamma$. 

The above output state coincides with the output state of the quantum switch in Eq. (\ref{outputswitch}). Hence, the circuit in Fig. 2 of the main text perfectly reproduces the output of the quantum switch of two constant channels.

\section{The coherently controlled scheme}
To identify the role of ICO in enabling anomalous heat flow, we analyze a thermalization process for a system
interacting with two identical thermalizing channels under coherent control. Specifically, 
when the control qubit is in state $\ket{0}$ ($\ket{1}$), the system interacts with thermalization channel 
$\mathcal{N}^{T_{\texttt{E}}}_{1}$ ($\mathcal{N}^{T_{\texttt{E}}}_{2}$).
The evolution map takes the form
\begin{equation}\label{map}
	\mathbb{M}_{ij} = 
	\left| 0 \right\rangle \left\langle 0 \right|_c \otimes K_1^j  
	+ \left| 1 \right\rangle \left\langle 1 \right|_c \otimes K_2^i ,
\end{equation}
where $K_{1(2)}^{i(j)}$ represent the Kraus operators for
the channels $\mathcal{N}_{1(2)}^{T_{\texttt{E}}}$ with the superscript 
$i(j)$ denoting the number of the operators.
By preparing the control qubit in the state
$\rho_{c}=\left|\psi\right\rangle \left\langle\psi\right|_c$, where
$\left|\psi\right\rangle_{c}=\sqrt{\alpha}\left|0\right\rangle_{c}+
\sqrt{1-\alpha}\left|1\right\rangle_{c}$, 
we create a coherent superposition of these two interactions.

\begin{figure}[b]
	\centering
	\begin{tabular}{cc}
	\includegraphics[scale=0.27]{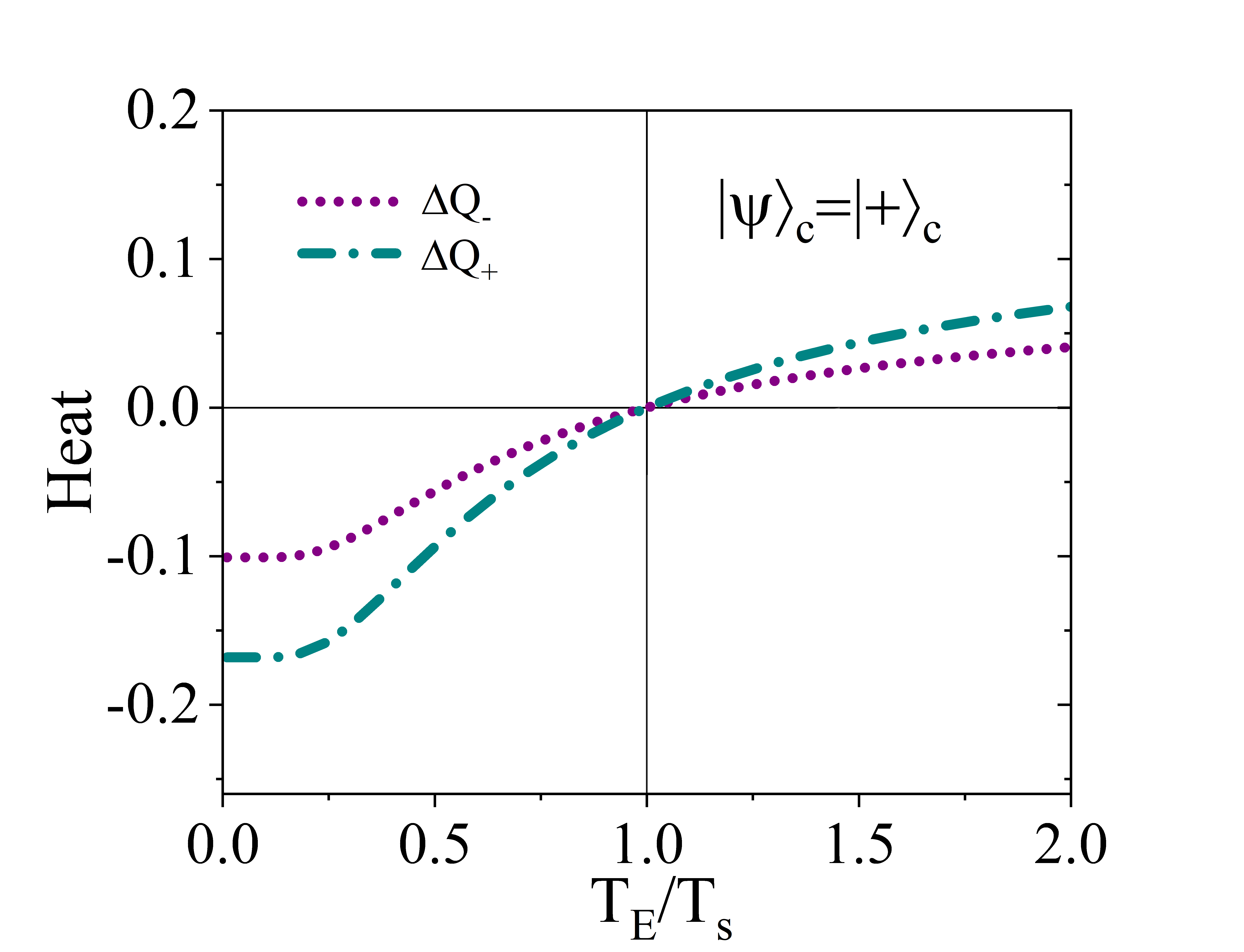}
	\end{tabular}
	\caption{Heat changes of the system, i.e., $\Delta Q^{\pm}$,
		against $T_{\texttt{E}}/T_{\texttt{S}}$, for coherently controlled processes.
		The control qubit with initial state $\ket{+}_{c}$ is finally measured in the
		$\ket \pm_{c}$ basis.  }\label{cohcontr}
\end{figure}

After making a measurement on the control qubit in the basis 
$\left\{\left|+\right\rangle_{c},\left|-\right\rangle_{c}\right\}$,
the system is projected into the corresponding conditional state $\rho_{\texttt{S}}^{\pm}$.
In Fig. \ref{cohcontr}, we present
the changes of the system's heat for each of the two possible measurement outcomes on the control qubit.
We see from the figure that for both measurement outcomes
no anomalous heat flow occurs: heat always flows from hot to cold.
In contrast, the ICO process described in the main text (see Fig. 4(b)) shows that anomalous heat flow can appear within a certain parameter window. Thus, the ICO process plays a crucial role in enabling anomalous heat flow.

\section{Experimental details}
\subsection{Heralded single-photon source}
\begin{figure}[h]
	\centering
	\begin{tabular}{cc}
		\includegraphics[scale=0.165]{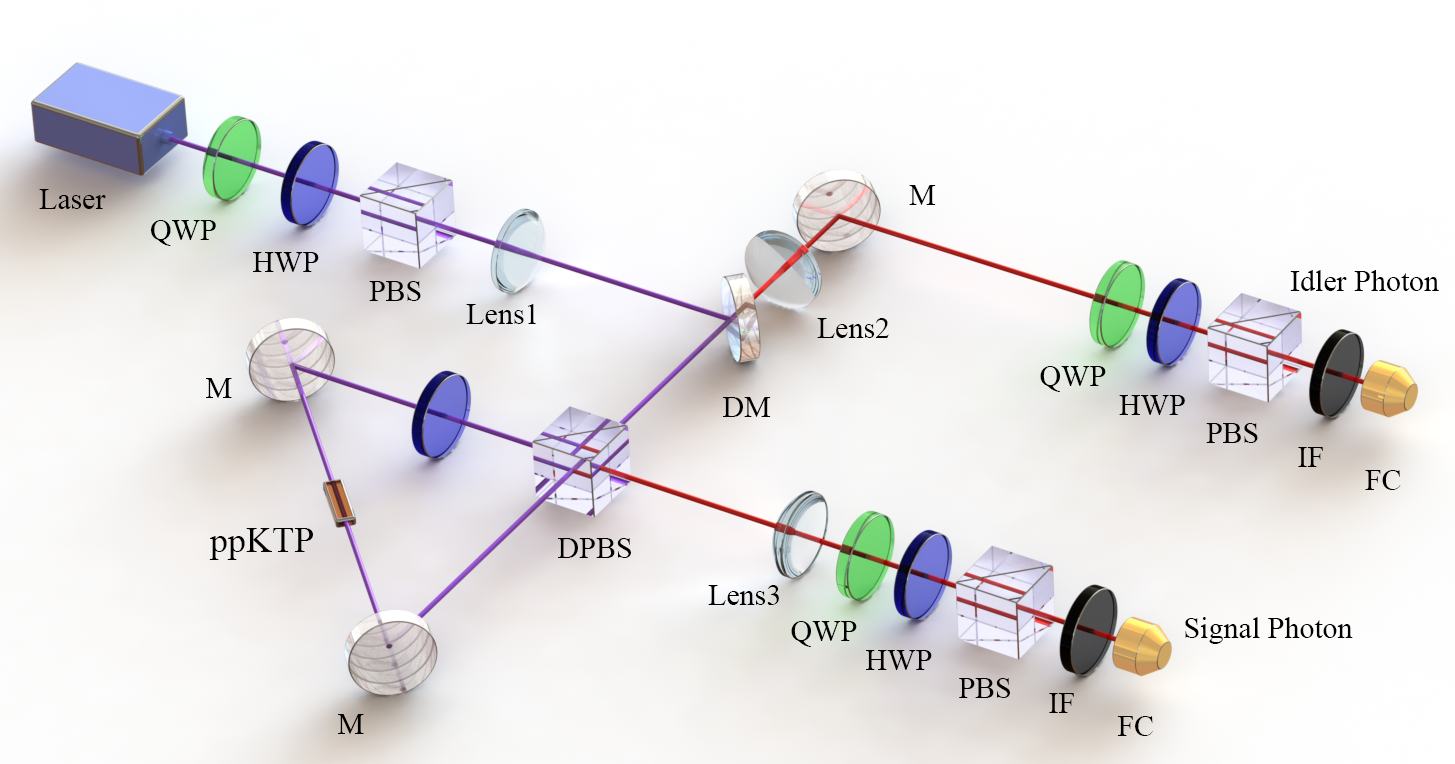}
	\end{tabular}
	\caption{Heralded single-photon source optical path diagram, quarter-wave plate (QWP), half-wave plate (HWP), polarization beam splitter (PBS), lens, dichroic mirror (DM), dual-wavelength polarization beam splitter (DPBS), dual-wavelength half-wave plate (DHWP), mirror (M), interference filter (IF), fiber collimator (FC).}
\end{figure}

A continuous-wave laser diode operating at 405 nm serves as the pump source. Its power is fine-tuned with a quarter-wave plate (QWP), half-wave plate (HWP) and polarizing beam-splitter (PBS) assembly, after which the beam is set to horizontal polarization. A 250 mm-focal-length plano-convex lens (lens 1) then focuses the beam to a 61 $\mu m$ waist on the input facet of the periodically poled potassium titanyl phosphate (ppKTP) crystal.
The ppKTP crystal is held in a temperature-stabilized oven at $15°C$ to satisfy type-II phase-matching, 
enabling efficient spontaneous parametric down-conversion (SPDC)
that yields polarization-entangled photon pairs centered at 810 nm.
With 1 $mW$ of pump power, we record approximately $5\times {{10}^{4}}$ coincidence counts per second
within a 3 $ns$ window.
A $45°$ dual-wavelength half-wave plate (DHWP) rotates the pump polarization so that a dual-polarizing beam splitter (DPBS) retro-reflects the residual pump.
Before the fiber collimator (FC), a 10 nm-bandwidth interference filter (IF) is employed to suppress background noise.
The idler photon is coupled into single-mode fiber (SMF) and detected directly by an avalanche photodiode, while the signal photon is delivered via SMF to the main optical path for subsequent experimental operations.

\subsection{The preparation the system's thermal state}
The preparation module of the system's initial thermal state 
is exhibited in FIG. 2 of the main text, which
can be described by the following procedure
\begin{eqnarray}
	&\left| H \right\rangle \xrightarrow{\text{HWP}@\theta }\cos 2\theta \left| H \right\rangle +\sin 2\theta \left| V \right\rangle &\nonumber\\
	&\xrightarrow{\text{unbalanced MZI}}{{\cos }^{2}}2\theta \left| H \right\rangle \langle H|+{{\sin }^{2}}2\theta \left| V \right\rangle \langle V|,&
\end{eqnarray}
where a photon in the horizontal polarization state
$\left| H \right\rangle$ passes through a HWP set at angle $\theta$, becoming a superposition of horizontal and vertical components, i.e., $\cos 2\theta \left| H \right\rangle +\sin 2\theta \left| V \right\rangle$,
and then enters an unbalanced Mach–Zehnder interferometer (MZI) composed of two PBSs and two mirrors.
Because the path difference between the two arms of the unbalanced MZI
exceeds the coherence length of the input state, the associated
coherence is destroyed and the state collapses into a thermal mixture.
We correspond the state $\left| H \right\rangle $ of the photon to the excited state $\left| e \right\rangle $ 
of the two-level system. The population $f_1$ of the system's excited state 
can thus be adjusted on demand by rotating the angle $\theta $ of HWP, 
with the relationship $f_{1}=\cos^{2} 2\theta$.

\subsection{Optical simulation of thermalization channels}
\begin{figure}[h]
	\centering
	\begin{tabular}{cc}
		\includegraphics[scale=0.25]{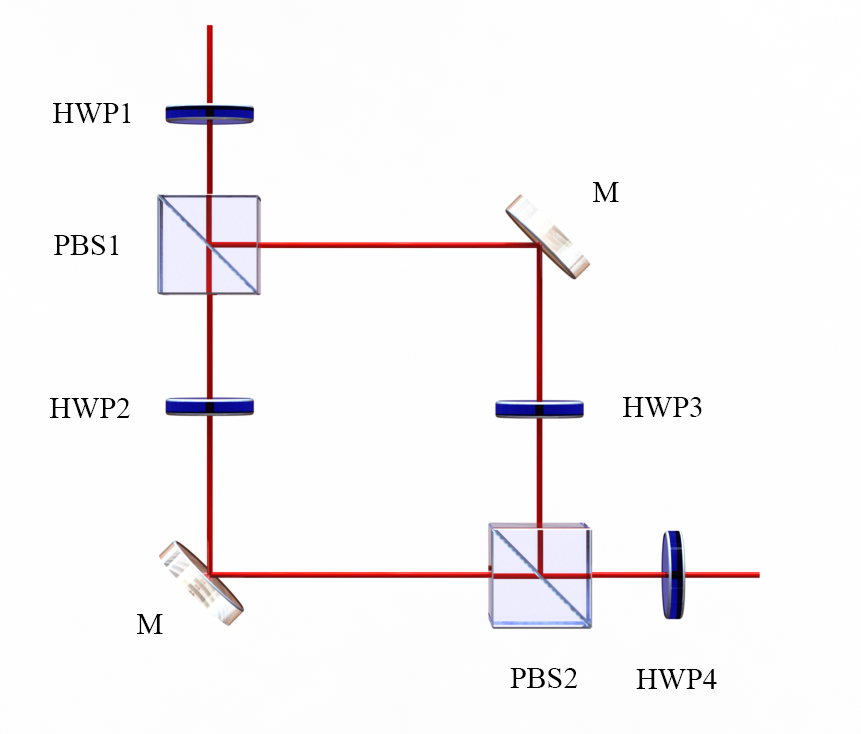}
	\end{tabular}
	\caption{Optical diagram of thermalization channel, half-wave plate (HWP), polarization beam splitter (PBS), mirror (M).}\label{Thercha}
\end{figure}

In the experiment, the Kraus operators in Eq. (\ref{eq4}) are implemented with the MZI
depicted in Fig. \ref{Thercha}, composed of four HWPs, two mirrors, 
and two PBSs.  The transformation of single-photon state
by HWP is shown in Eq. (\ref{HWP}).
The correspondence
between the Kraus operators and the angle settings of four HWPs is listed in Table \ref{tab1}.
\begin{table}[h]
	\centering
	\begin{tabular}{|c|c|c|c|c|}
		\hline
		Kraus operator & HWP1 & HWP2 & HWP3 & HWP4 \\
		\hline
		$K_{1(2)}^{0}=\sqrt{p}\left| e \right\rangle \left\langle  e \right|$ & 0 & 0 & $\frac{\pi}{4}$ & 0 \\
		\hline
		$K_{1(2)}^{1}=\sqrt{p}\left| e \right\rangle \left\langle  g \right|$ & $\frac{\pi}{4}$ & 0 & $\frac{\pi}{4}$ &0 \\
		\hline
		$K_{1(2)}^{2}=\sqrt{1-p}\left| g \right\rangle \left\langle  g \right|$ & 0 & $\frac{\pi}{4}$ & 0 & 0 \\
		\hline
		$K_{1(2)}^{3}=\sqrt{1-p}\left| g \right\rangle \left\langle  e \right|$ & $\frac{\pi}{4}$ & $\frac{\pi}{4}$ & 0 & 0 \\
		\hline
	\end{tabular}
	\caption{The correspondence table between the Kraus operators of the thermal channel and the angles of the half-wave plate set $\left\{ \text{HWP}1,\text{HWP}2,\text{HWP}3,\text{HWP}4 \right\}$.}\label{tab1}
\end{table}

We take the experimental realization of $K_{1(2)}^{0}$ as an example and
set the angles of HWP group $\left\{ \text{HWP}1,\text{HWP}2,\text{HWP}3,\text{HWP}4 \right\}$
to $\left\{ 0,0,\frac{\pi}{4},0 \right\}$. 
In this case, the MZI leaves the horizontal component $a\left| H \right\rangle $ of any single-photon input state $\left| \psi  \right\rangle =a\left| H \right\rangle +b\left| V \right\rangle $ unchanged, while discarding the vertical component $b\left| V \right\rangle$, which is the effect achieved by the operator $K_{1(2)}^{0}$.
In correspondence with the four operators, we conduct four sets of independent experiments and construct the
associated evolved states of the system.
The final evolved state of the system through the thermalization channel can be
obtained by multiplying the acquired data by the parameters $p$ or $1-p$
in post-processing.

\begin{figure*}[t]
	\centering
	\begin{tabular}{cc}
		\includegraphics[scale=0.42]{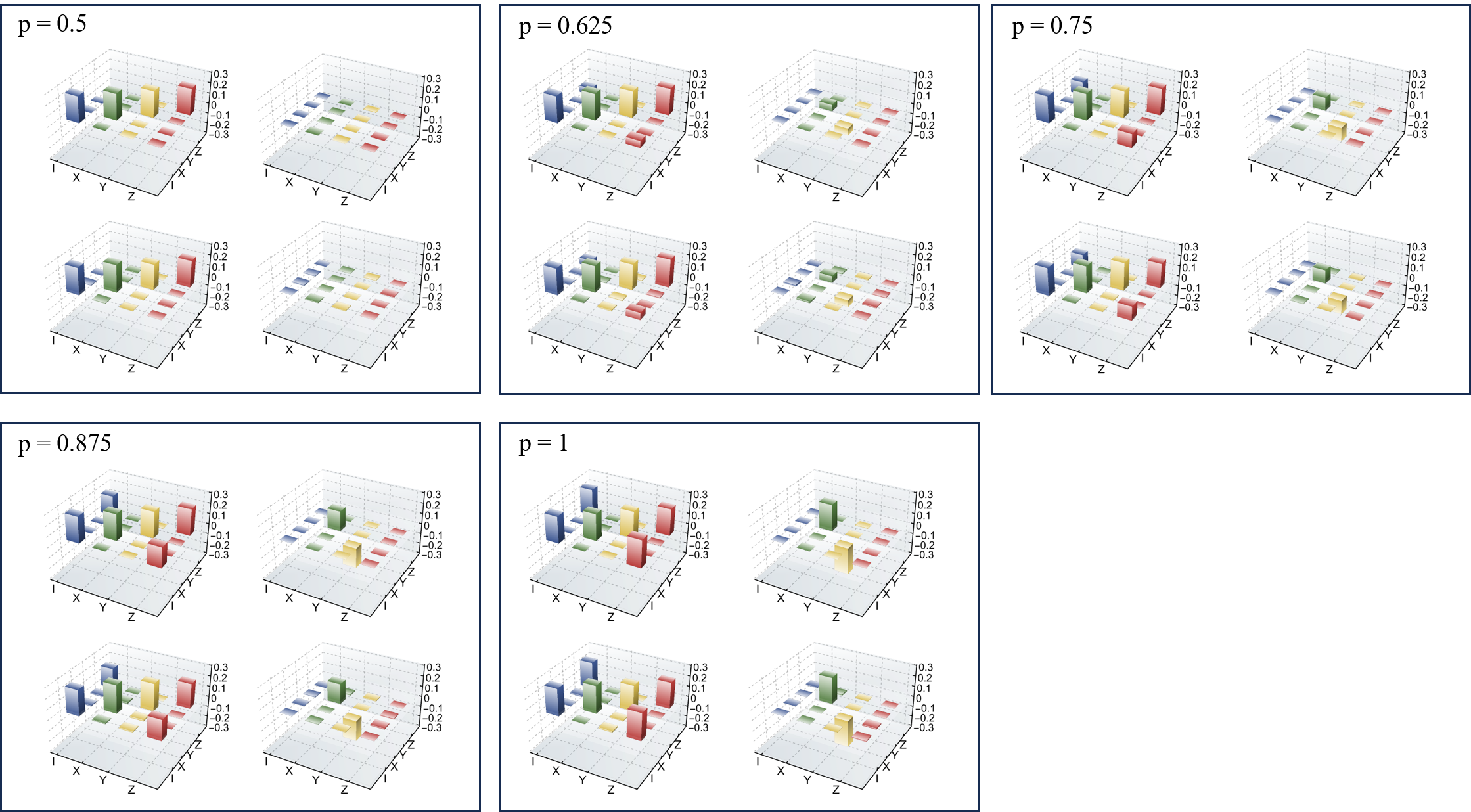}
	\end{tabular}
	\caption{Set the thermalization channel parameter $p=\left\{ 0.5,0.625,0.75,0.875,1 \right\}$. The single-photon quantum process tomography results under the Pauli basis are presented as follows: Within each rectangular box, the top two plots display the real part (left) and imaginary part (right) of the theoretical process matrix, while the bottom two plots show the real part (left) and imaginary part (right) of the experimentally reconstructed process matrix.}\label{procetom}
\end{figure*}

We perform single-photon quantum process tomography to characterize the thermalization channel.  
We sequentially prepare the input states $\{ \left| H \right\rangle, \left| V \right\rangle,
\left| D \right\rangle =\frac{1}{\sqrt{2}}\left( \left| H \right\rangle +\left| V \right\rangle  \right), 
\left| R \right\rangle =\frac{1}{\sqrt{2}}\left( \left| H \right\rangle -i\left| V \right\rangle  \right) \}$
of the single photons, let them evolve through the thermalization channel, and reconstruct the output states via Pauli-basis
measurements $\left\{ {{\sigma }_{x}},{{\sigma }_{y}},{{\sigma }_{z}} \right\}$.
Repeating the procedure for $p=\left\{ 0.5, 0.625, 0.75, 0.875, 1 \right\}$ yields the 
process matrices shown in Fig. \ref{procetom}. The corresponding fidelities, i.e., 
$0.99954\pm0.00021$, $0.99853\pm0.00016$, $0.99887\pm0.00014$, $0.99706\pm0.00029$, and $0.99808\pm0.00063$, 
confirm the high-fidelity implementation of the thermalization channel.

\subsection{Experimental implementation of quantum switch}
Our optical quantum switch is realized via a folded MZI approximately 1.8 meters
in length, as shown in Fig.2 of the main text. 
BS1 bifurcates the photon into two spatial modes that serve as a control qubit:
one arm undergoes the causal order $\mathcal{N}_{1}^{T_{e}}\circ \mathcal{N}_{2}^{T_{e}}$, the other $\mathcal{N}_{2}^{T_{e}}\circ \mathcal{N}_{1}^{T_{e}}$. 
BS2 recombines the modes coherently, 
projecting the control qubit onto the ${{\left| \pm \right\rangle }_{c}} $ basis.

To precisely match the optical path lengths of the two interferometer arms,
a platform adjustment mount holding two right-angle prism mirrors is placed on a precision
translation stage in one arm of the interferometer.
By adjusting the position of the translation stage, the optical path difference between the two arms
can be effectively compensated.

To suppress external noise and guarantee long-term stability, we adopted an active phase-locking system.
A single-longitudinal-mode laser with a central wavelength of 808 nm injects a horizontally polarized reference beam
backward into the folded MZI. 
This beam is vertically offset by approximately 5.5 mm from the signal beam and further isolated by a 3 nm-bandwidth interference filter (IF) in front of the fiber coupler (FC). 
Owing to this separation, the reference light bypasses all HWPs inside the interferometer, 
preserving its initial polarization.
After traversing the MZI, the reference beam is coupled into a single-mode fiber (SMF)
and monitored with a silicon photodetector (PDA100A2).
The resulting electrical signal is sent to a servo control system (STEMlab125-14). 
The output signal from the servo system is amplified twice (using an HMC580 amplifier and a high-voltage amplifier, HVA200) before being fed into a piezoelectric transducer (PZT). 
Through negative feedback, the phase is ultimately stabilized at the preset value.

We now explain how to realize Kraus operators $\mathbb{W}{ij}$, Eq. (\ref{eq5}),
of the quantum switch by tuning the angles of HWPs, i.e.,
$\{\text{HWP}1,\text{HWP}2,\text{HWP}3,\text{HWP}4\}$,
within the two MZIs (cf. Fig. \ref{Thercha} and Fig. 2 of the main text)
that implement the thermalizing channels $\mathcal{N}_{1}^{T{e}}$ and $\mathcal{N}_{2}^{T{e}}$.
Taking the operator $\mathbb{W}_{13}=
\left| 0 \right\rangle_c \left\langle 0 \right| \otimes K_1^1 K_2^3 
+ \left| 1 \right\rangle_c \left\langle 1 \right| \otimes K_2^3 K_1^1$ as an example,
we need to set the angles of HWPs corresponding to the operator $K_1^1$
of $\mathcal{N}_{1}^{T_{e}}$ to $\{\pi/4,0,\pi/4,0\}$,
and the angles of HWPs corresponding to the operator $K_2^3$ 
of $\mathcal{N}_{2}^{T_{e}}$ to $\{\pi/4,\pi/4,0,0\}$.
This setting makes the combined effect of $K_1^1 K_2^3$ to be a projection
operation in terms of $\left|e\right\rangle\left\langle e\right|$
on the system, while the combined effect of $ K_2^3 K_1^1$ is an operation of
$\left|g\right\rangle\left\langle g\right|$.
By conducting 16 separate experiments, each corresponding to the action of a Kraus operator
of the quantum switch, we obtain 16 evolved states of the system through quantum state tomography.
In the post-processing of the data, we multiply all the evolved states by the corresponding
channel parameters associated with 
$p$ and sum them up,
which eventually allows us to construct the final state $\rho _{s}^{\pm }$ of the system after the quantum switch.

\subsection{The simulation of system's dynamics in work strokes of Otto cycle}
\begin{figure}[h]
	\centering
	\begin{tabular}{cc}
		\includegraphics[scale=0.2]{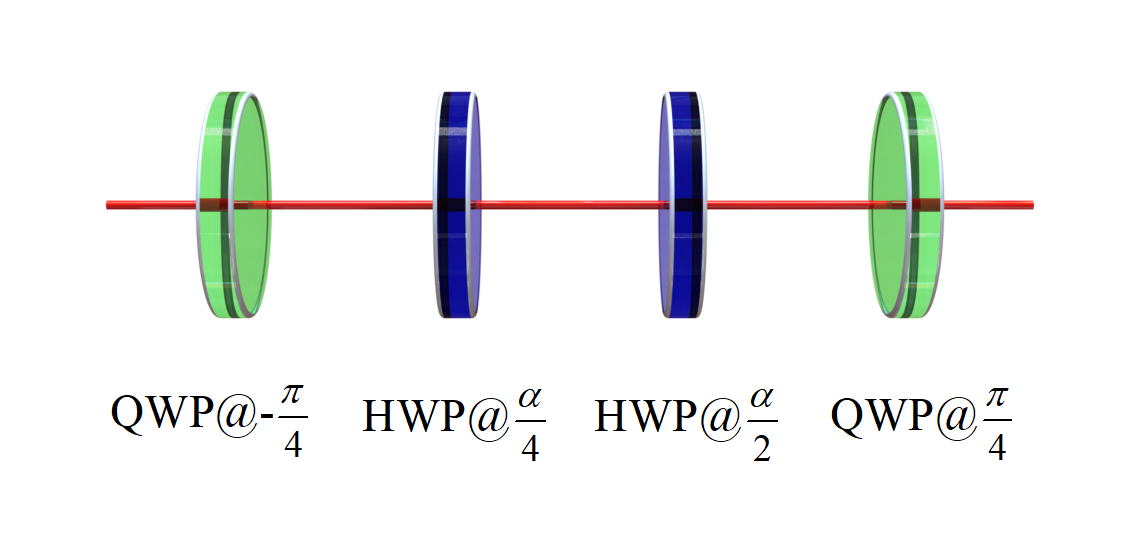}
	\end{tabular}
	\caption{Optical Path Schematic for Adiabatic compression/expansion process, quarter-wave plate (QWP), half-wave plate (HWP).}\label{Jones}
\end{figure}

In both quantum adiabatic compression and expansion strokes, the system experiences a
unitary dynamics driven by time-dependent Hamiltonians, given as
$H_{c}(t)=\omega_{c}(t)\sigma_{z}/2$ and $H_{e}(t)=\omega_{e}(t)\sigma_{z}/2$, respectively.
Specifically, we set $\omega _{c}(t)=\omega_{1}\left(1-\frac{t}{\tau}\right)+\omega_{2}(\frac{t}{\tau })$
and $\omega_{e}(t) = \omega_{2} \left(1 - \frac{t}{\tau}\right)+\omega_{1}(\frac{t}{\tau})$ with $\tau$
the duration of each stroke, which
ensure that in the compression stoke the system's frequency changes from $\omega_{1}$
to $\omega_{2}>\omega_{1}$, while in the expansion stoke the frequency changes back from
$\omega_{2}$ to $\omega_{1}$. 
The system's dynamics in both
adiabatic processes can be described by the unitary operator
\begin{equation}
U_{c(e)}(t)=\exp \left[-i\int_{0}^{t}H_{c(e)}(t^{\prime})dt^{\prime}\right].
\end{equation}
By setting $\omega^{\prime}_{c(e)}  =t^{-1}\int_{0}^{t}\frac{\omega_{c(e)}(t^{\prime})}{2}dt^{\prime}$, the time-evolution operator can be rewritten as
\begin{equation}\label{Ucet}
	U_{c(e)}(t)=\exp [-i\omega^{\prime}_{c(e)} t \sigma_{z}].
\end{equation}

In order to simulate the system's dynamics in experiment, we 
adopt the polarization rotation Jones matrix defined as  
\begin{equation}\label{RZ}
	{{R}_{Z}}\left( \alpha  \right)=\exp [-i\frac{\alpha }{2}{{\sigma }_{z}}]=\left( \begin{matrix}
		{{e}^{\frac{-i\alpha }{2}}} & 0  \\
		0 & {{e}^{^{\frac{i\alpha }{2}}}}  \\
	\end{matrix} \right)
\end{equation}
Eqs. (\ref{Ucet}) and (\ref{RZ}) show that a correspondence between evolution operator
$U_{c(e)}(t)$ and the Jones matrix ${{R}_{Z}}\left( \alpha_{c(e)}  \right)$ can be established by taking
\begin{equation}
	\alpha_{c(e)} =2\omega^{\prime}_{c(e)} t=\int_{0}^{t}{\omega_{c(e)} (t^{\prime}) }dt^{\prime}.
\end{equation}
For our specific forms of $\omega_{c(e)} (t)$ and duration $\tau$ of the dynamics, 
we can obtain that $\alpha_{c}=\alpha_{e}=(\omega_{1}+\omega_{2})\tau/2$.
This means that, by manipulating the parameter $\alpha_{c(e)}$ in the Jones matrix,
we can simulate the system's dynamics driven by $H_{c(e)}(t)$ with given 
$\omega_{1}$, $\omega_{2}$ and $\tau$ in adiabatic compression (expansion) stroke.
In the experiment, as shown in Fig.~\ref{Jones}, the Jones matrix described by Eq.~(\ref{RZ}) can be
implemented through a combination of QWPs and HWPs.
The matrix representations of the HWP and QWP read
\begin{equation}\label{HWP}
	\text{HWP}@\theta=\left( \begin{matrix}
		\cos 2\theta & \sin 2\theta  \\
		\sin2\theta & -\cos2\theta  \\
	\end{matrix} \right), 
\end{equation}
and
\begin{equation}
	\text{QWP}@\theta=\left( \begin{matrix}
	\cos^2\theta+i \sin^2\theta &  (1-i)\sin\theta\cos\theta \\
		(1-i)\sin\theta\cos\theta & \sin^2\theta+i \cos^2\theta  \\
	\end{matrix} \right),
\end{equation} 
respectively. Thus, we have
\begin{eqnarray}
	{{R}_{Z}}\left( \alpha  \right)&=&\text{QWP}@\frac{\pi }{4}\cdot \text{HWP}@\frac{\alpha }{2}\cdot \text{HWP}@\frac{\alpha }{4}\cdot \text{QWP}@-\frac{\pi }{4}\nonumber\\
	&=&\left( \begin{matrix}
		{{e}^{-\frac{i\alpha }{2}}} & 0  \\
		0 & {{e}^{\frac{i\alpha }{2}}}  \\
	\end{matrix} \right).
\end{eqnarray} 
Taking the adiabatic compression stroke as an example, we explain
the correspondence between the rotation angles of the wave plates and the frequency
changes of the system. We set $\omega_1 \tau = \pi$ and consider the case where
$\omega_2 = 1.5 \omega_1$. According to the adiabatic compression Jones parameter
$\alpha_c = (\omega_1 + \omega_2) \tau / 2$, the corresponding angles for HWP$@\alpha_c/4$
and HWP$@\alpha_c/2$ are $1.25\pi/4$ and $1.25\pi/2$, respectively.

\end{document}